\documentclass[conference]{IEEEtran}
\makeatletter
\def\ps@headings{%
\def\@oddhead{\mbox{}\scriptsize\rightmark \hfil \thepage}%
\def\@evenhead{\scriptsize\thepage \hfil \leftmark\mbox{}}%
\def\@oddfoot{}%
\def\@evenfoot{}}
\makeatother
\usepackage{graphicx}
\usepackage{times}
\usepackage{cite}
\usepackage[english]{babel}
\usepackage[latin1]{inputenc}
\usepackage[font=small,format=plain,labelfont=bf,up,textfont=it,up]{caption}

\usepackage{amsmath}
\usepackage{amssymb}
\usepackage{algorithmic}
\usepackage{algorithm}

\newcommand{\PP}[1]{\mathrm{P}\{ #1 \}}
\newcommand{\cP}[2]{\mathrm{P}\{ #1 \mid #2 \}}
\newcommand{\E}[1]{\mathrm{E}[ #1 ]}
\newcommand{\cE}[2]{\mathrm{E}[ #1 \mid #2 ]}

\begin{document}

\title{Criticality of Large Delay Tolerant Networks via 
  Directed Continuum Percolation in Space-Time}
\author{%
\IEEEauthorblockN{Esa Hyyti{\"a} and J{\"o}rg Ott}
\IEEEauthorblockA{Department of Communications and Networking\\Aalto University, Finland}}
\maketitle

\begin{abstract}
We study delay tolerant networking (DTN) and in particular,
its capacity to store, carry and forward messages 
so that the messages eventually reach their final destination(s).
We approach this broad question in the framework
of percolation theory. To this end, we assume an elementary
mobility model, where nodes arrive to an infinite plane
according to a Poisson point process,
move a certain distance $\boldsymbol{\ell}$, and then depart. 
In this setting, we characterize the mean density of nodes 
required to support DTN style networking.
In particular, under the given assumptions, we show that
DTN communication is feasible when the mean node degree $\boldsymbol{\nu}$
is greater than  $\mathbf{4\cdot\boldsymbol{\eta}_c(\boldsymbol{\gamma})}$, 
where parameter $\boldsymbol{\gamma}\mathbf{=\boldsymbol{\ell}/d}$
is the ratio of the distance $\boldsymbol{\ell}$ to the transmission range $\mathbf{d}$,
and $\mathbf{\boldsymbol{\eta}_c(\boldsymbol{\gamma})}$ 
is the critical reduced number density of tilted cylinders
in a directed continuum percolation model.
By means of Monte Carlo simulations, we give numerical
values for $\mathbf{\boldsymbol{\eta}_c(\boldsymbol{\gamma})}$.
The asymptotic behavior of $\mathbf{\boldsymbol{\eta}_c(\boldsymbol{\gamma})}$
when $\boldsymbol{\gamma}$ tends to $\boldsymbol{\infty}$
is also derived from a fluid flow analysis.
\end{abstract}

\section{Introduction}

Delay-tolerant networking (DTN) is a last-resort networking paradigm for mobile nodes when
direct and multi-hop connections are infeasible, i.e.,
the network is not connected and nodes have to \emph{carry}
the messages to the next node (store-carry-forward).
We study the capacity of DTN to deliver a message to its destination(s)
in the framework of the percolation theory. 
To this end, we assume an elementary
mobility model where nodes arrive according
to a Poisson point process on a plane, move a certain distance $\ell$,
and then depart. 
Nodes perform epidemic routing, i.e., whenever two nodes
meet, all messages are exchanged.
We are interested in
finding the circumstances under which the lifetime of a message 
becomes infinite (with some positive probability)
so that
the message could reach a recipient located
at an arbitrary distance from the source.
We obtain a fundamental criticality condition that
characterizes the sufficient mean density of nodes
to this end.
The criticality condition
takes form $\nu > \nu_c(\gamma) = 4\,\eta_c(\gamma)$, 
where $\nu$ denotes the \emph{mean node degree} (number of neighbors),
$\gamma$ is the ratio of $\ell$ to the transmission range $d$,
and $\eta_c(\gamma)$ is an unknown function, which we determine
in this paper.

For a mobile DTN, we find that $\nu_c(\gamma) \le 1.52$ for all $\gamma$,
while only at $\nu \approx 4.51$ a (non-DTN) wireless ad-hoc network 
percolates and a gigantic connected cluster emerges
\cite{franceschetti-info-theory-2007}.
In other words, DTN communication is possible over
a very sparse network provided that the network's topology changes
in time.
A convenient characterization in fact is to say
that \emph{DTN is a network that is super-critical in space-time.}
In practice, DTN is often sub-critical at a random time instant and
the messages find their way to the destination through the store-carry-forward
routing in space-time.

This work is motivated by different opportunistic networking schemes.
One such scheme is \textit{Floating Content}, where nodes replicate
messages only within the area where each message is deemed relevant.
A criticality condition,
characterizing the circumstances under which the lifetime
of the message can be expected to be long,
was established in \cite{hyytia-infocom-2011} under the assumptions
of mobile nodes and point contacts (i.e., the transmission range is small compared
to the dimension of the area). 
Similarly, Beachnet \cite{ott-extreme-2011}
the aim is to produce quasi-periodic ``information waves'' carried by an
underlying field of immobile nodes (say, devices on a beach).
Beachnet is designed to offer regular content dissemination when
nodes are (mostly) stationary and
their collective cooperation, based on simple local transmission rules, is sought.
Similar concepts for one-to-many content dissemination include hovering information \cite{villalba-tr-2007}
and ad-hoc podcasting \cite{lenders-sigmobile-2008},
but our model naturally covers the special case of one-to-one messaging.
These types of DTN schemes introduce many interesting problems.
Perhaps the most fundamental question is whether
a network can transport messages to their intended destinations or not,
and this is also our focus in this paper.

\subsection{Background and related work}

We assume epidemic routing which was proposed 
by Vahdat and Becker in \cite{vahdat-tr-2000}.
The operational principle is very simple:
whenever two nodes meet they exchange the messages only one of them has.
The performance analysis of the epidemic routing often assumes
exponentially distributed (i.i.d.)  inter-meeting times
leading to Markovian models \cite{groenevelt-peva-2005}.
As the size of the network grows, solving Markov chains becomes
infeasible, and Zhang et al., in \cite{zhang-compnet-2007}
obtained ODEs as a fluid limit of such Markovian models.
In above work, the spatial dimension has been abstracted away.

Jacquet et al.~\cite{jacquet-info-2010} 
consider the propagation speed of the information in DTN
for a fixed set of nodes moving in a finite region.
The spatial dimension is explicitly present in their formulation.
Somewhat related,
Grossglauser and Tse \cite{Grossglauser-Networking-2002} studied
the performance of adhoc networks under mobility.
They focused on connected networks, where nodes can either communicate directly
or via multi-hop connections, and showed that mobility increases the overall
capacity in the network given the users tolerate additional delays.

In this paper, we consider a fundamental feasibility problem of DTN-style
communication with the elementary mobility model
by means of percolation theory \cite{stauffer-1994,meester-1996}.
Percolation theory has been succesfully applied to study the performance
of wireless multi-hop networks since the early work by Gilbert \cite{gilbert-siam-1961}.
Similarly as in \cite{gilbert-siam-1961}, most of the work utilizes
undirected planar continuum percolation models to argue, e.g., 
about the network's connectivity or capacity. 

In contrast, 
in \cite{hyytia-cl-2012}, we assumed \emph{stationary nodes} and showed
that opportunistic content dissemination schemes, such as the floating content,
can be analyzed by using a three-dimensional continuum percolation model.
In this paper, we adapt the same approach, but instead of approximating
the process by undirected percolation model, we consider 
the \emph{actual directed percolation}
characterizing the dissemination of messages by \emph{mobile nodes}
exactly in a DTN network.
There are less results available for the directed percolation models than
there are for the normal undirected percolation, and
results for directed continuum models are even more scarce.
The basic directed cases are the bond and site percolation models,
where, e.g., bonds have fixed directions (e.g., up and right in a square lattice)
\cite{duarte-zeitschrift-1990,stauffer-1994}.
Directed models have been also studied in the context of scale-free 
networks \cite{schwartz-physreve-2002},
which arise in Internet, social networks etc.

The direction property can be defined many ways.
The model considered in this paper, is a specific case of directed
continuum percolation, where three-dimensional
objects may belong to a given cluster fully or partially.
In particular, %
the third dimension is time and, 
due to the causality, only the post-contact part of an object
belongs to the given cluster.
In this setting, we obtain an elegant fundamental result 
for the minimum feasible mean node degree to support DTN communication
as a function of the ratio of movement to the transmission range.
This critical percolation threshold is determined by Monte Carlo simulations.
The obtained results define the feasible operation regime for large DTN networks.

The rest of the paper is organized as follows. 
In Section~\ref{sect:notation} we introduce the model and the notation.
Section~\ref{sect:analysis} contains the analysis and
the description of the Monte Carlo simulations.
The numerical results are given in Section~\ref{sect:results},
and Section~\ref{sect:conclusions} concludes the paper.

\section{Model and Notations}
\label{sect:notation}

In this section, we describe our model, including the 
arrival process, nodes' mobility and the assumptions regarding
the radio communication. %
Throughout this paper, we implicitly assume a large network
and a long distance, either in space and/or time, between
the source and the destination.

\subsubsection{Node mobility}
We assume that nodes arrive according to a Poisson point process on an infinite
plane with rate density denoted by $\lambda$ [node/m$^2$/s]. 
Mobility is specified with a constant movement $\ell$, i.e.,
during the constant time $t$ 
a node moves the distance of $\ell$
to a random direction (isotropic). 
After the movement, the node departs from the system.
Note that this activity cycle from ``an arrival'' to ``a departure''
can model an activity period of a node (from active state to inactive state),
or the departure can correspond to an event where a node 
replaces the information content with another.

The node density in plane is
$$
\tilde{n} \triangleq \lambda\cdot t,
$$
The arrival-departure cycle can be interpreted as the time
when a mobile node is participating in the DTN activity.
In the basic stationary case, $\ell=0$
and the nodes remain still for the constant time $t$ before departing.
We study the capacity of a large network to transport messages 
in space-time. Thus,
the node mobility model applies to the intermediate nodes, while
the source and destination node can be thought to be, e.g., 
two permanent nodes located far apart from each other.

\subsubsection{Information dissemination}
Whenever two nodes are within each others' transmission range $d$, i.e.,
the distance between them is less than $d$, they immediately exchange
the messages. %
The assumption of a fixed transmission range
is often referred to as the Gilbert's disc model or boolean model
\cite{gilbert-siam-1961}.
The mean number of neighbors a node has, i.e., the mean node degree
is thus $\nu = \tilde{n}\cdot \pi d^2$.
In the percolation theory, two objects are connected if they overlap
and thus the corresponding radius $r$ is half of the transmission range,
$$ r = d/2.$$

\begin{figure}
  \centering
  \includegraphics[width=40mm]{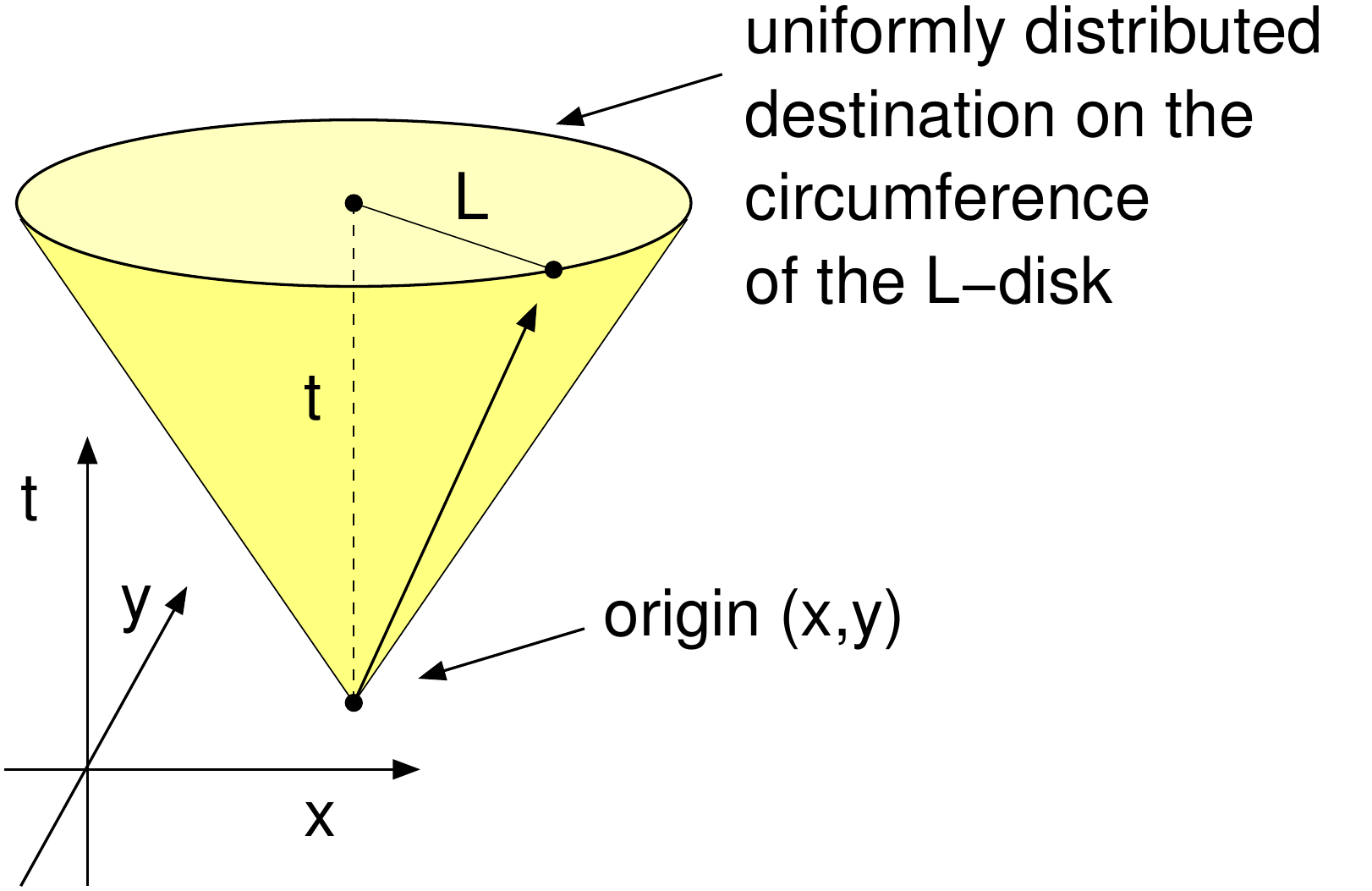}
  \hspace{-10mm}
  \includegraphics[width=40mm]{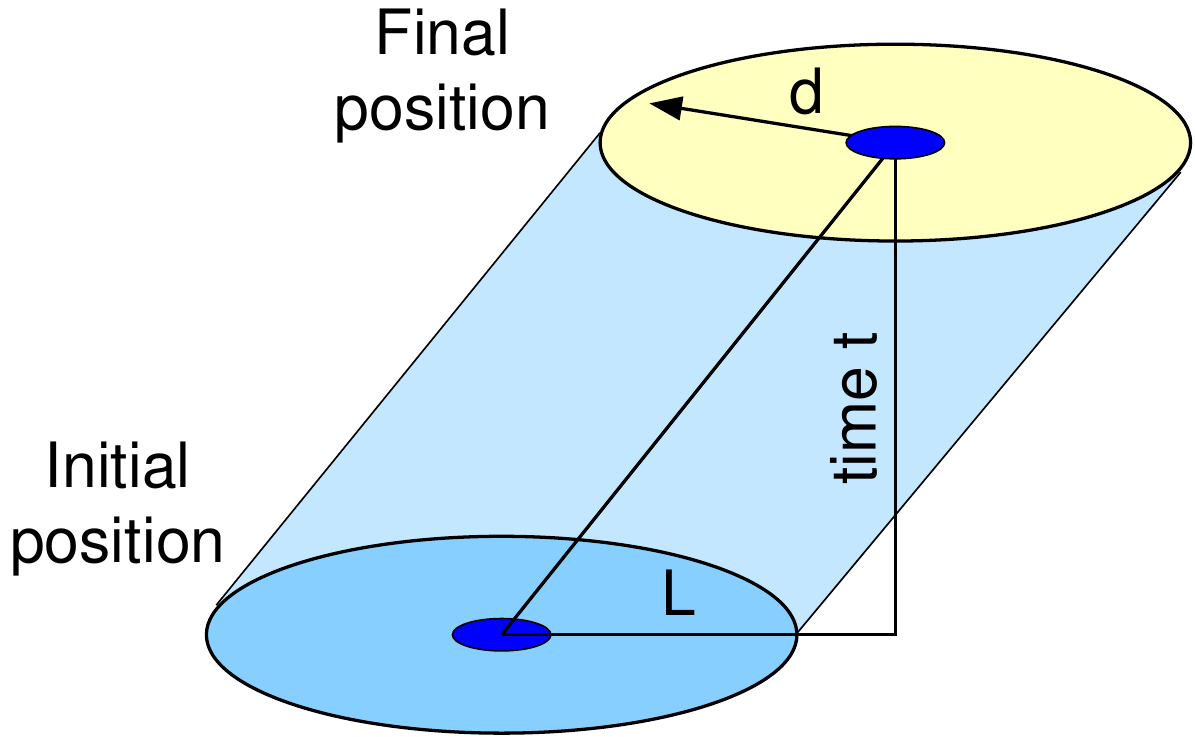}
  \caption{Node mobility in space-time: (left) during time $t$, a node
    moves a distance of $\ell=L$ to a random direction in $(x,y)$-plane
    and departs. (right) The volume covered corresponds to a tilted cylinder.}
  \label{fig:movement}
\end{figure}

\subsubsection{Space-time}
The movement and the radio coverage of a node in
space-time, comprising $(x,y)$-plane and the time axis, 
are depicted in Fig.~\ref{fig:movement}.
The radio coverage at any time instant is a disc.
As nodes move in time,
the coverage in 3D \emph{space-time} is a sheared cylinder.
The \emph{tilt} of the cylinder is defined 
by the ratio of the movement to the transmission range, 
$$
\gamma \triangleq \frac{\ell}{d} = \frac{\ell}{2r}.
$$
Somewhat counter-intuitively,
the number density of nodes (cylinders) in \emph{space-time}
is equal to the arrival rate $\lambda$,\footnote{Consider a volume $V=AT$, with area $A$ in $(x,y)$-plane and
height $T$. The mean number of nodes in $V$ is $\lambda V$, i.e., the node density $n$ %
is $\lambda$.}
$$
 n = \lambda.
$$
Instead of number density, the density of objects is
often expressed using
either the \emph{reduced number density} $\eta$,
$$
\eta \triangleq n \cdot V,
$$
where $V$ is the volume of the shape,
or the \emph{volume fraction} $\phi$, for which it holds that
$\phi = 1 - e^{-\eta}$.

In our case, $V=\pi r^2\cdot t$ independently of $\gamma$.
Without lack of generality, we can scale the space-time.
The time axis can be scaled by constant $\alpha$,
e.g., so that $t=1$.
Similarly, the $(x,y)$-plane can be scaled by constant $\beta$,
e.g., so that $2\cdot\ell+d=1$.
The ratio $\gamma$ is clearly invariant under such scalings.
Similarly,
$$ n \propto \alpha^{-1}\beta^{-2}
\quad
\text{and}
\quad
V \propto \alpha \beta^2,
$$
and thus also the reduced number density $\eta=n\cdot V$ and
the volume fraction $\phi$ are invariant.

\section{Analysis}
\label{sect:analysis}

In this section, we first study the corresponding directed
continuum percolation model characterizing the evolution of a
message in space-time. Then we describe Monte Carlo simulations
to estimate the critical percolation threshold, and also
derive an asymptotic scaling law for the threshold.

\subsection{Directed continuum percolation}

We consider a dynamic system in space-time, where each node corresponds
to a (tilted) cylinder as depicted in Fig.~\ref{fig:movement}.
Two nodes can communicate from the moment 
the distance between them becomes less than the constant transmission range $d$.
It follows that the process describing the evolution of 
a message in the
infinite plane is equivalent to a \emph{directed continuum percolation}
of aligned or sheared cylinders in three dimensions \cite{hyytia-cl-2012}.
The radius $r$ of the cylinders with respect to percolation is equal to
$d/2$, and the constant holding time $t$ corresponds to the height of the cylinders.
The moment two $r$-cylinders touch each other, messages can be transmitted.
The causality imposes that the information flows only in the direction of the positive
time axis.

Formally, let $a \leadsto b$ denote that a path from node $a$ to node $b$
exists in space-time in the sense that $a$ could send a message to $b$ 
(potentially via a multi-hop connection).
Due to the causality, $\leadsto$ is not a symmetric relation,
$$
a \leadsto b \quad \nRightarrow \quad b \leadsto a.
$$
Moreover, $\leadsto$ is not even transitive,
$$
a \leadsto b \text{ and } b \leadsto c \quad \nRightarrow \quad a \leadsto c,
$$
as a message from $a$ to $b$ may arrive too late for $b$
to deliver it to $c$.
Let $\mathcal{S}(a)$ denote the set of nodes
reachable from $a$,
$$
\mathcal{S}(a) = \{ b \,:\, a \leadsto b \}.
$$
We are interested in the size (i.e., the cardinality)
of $\mathcal{S}(a)$, and let random variable $S$, \emph{cluster size},
denote this quantity,
$$
S \triangleq |\mathcal{S}(a)|.
$$

\begin{figure}
  \centering
  \begin{tabular}{c@{\hspace{0mm}}c@{\hspace{0mm}}c}
    \includegraphics[clip,trim=160 0 295 0,height=40mm]{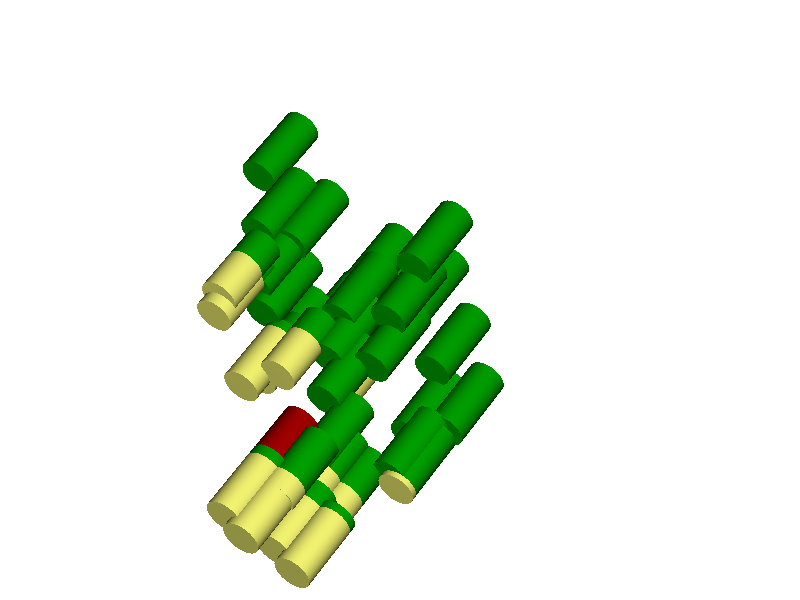} & 
    \includegraphics[clip,trim=160 0 340 0,height=40mm]{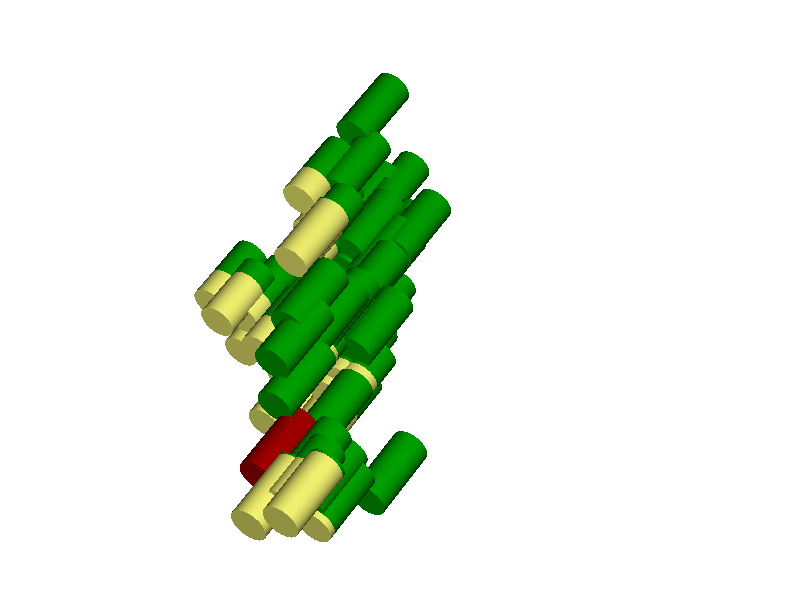} & 
    \includegraphics[clip,trim=190 0  70 0,height=40mm]{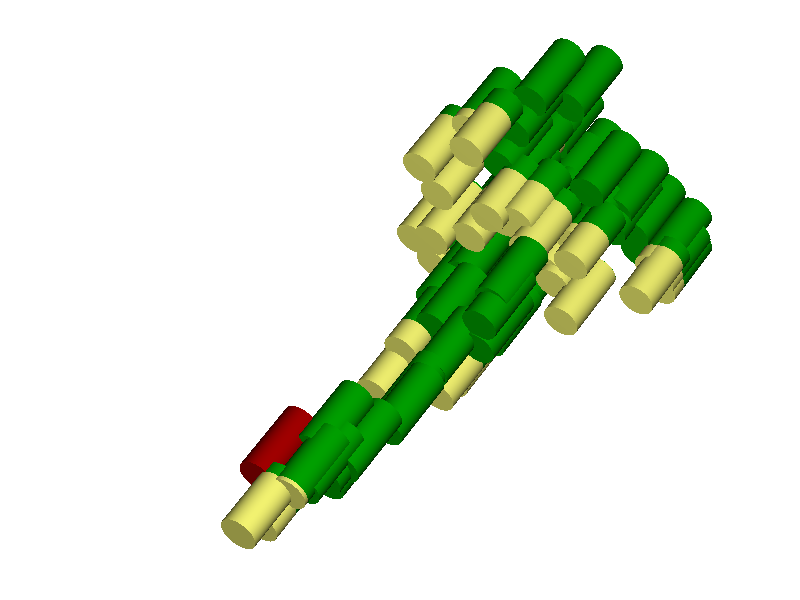} \\
    $37$ nodes &
    $67$ nodes &
    $86$ nodes
  \end{tabular}
  \caption{Sample realizations of dissemination of a message with
    stationary nodes. 
    The reduced number density is $\eta=0.37$. %
    The red cylinder is the source,
    the light yellow color indicates that a node is waiting for the message
    and the green color that a node has acquired it.
    Nodes which did not acquire the message have been omitted.}
  \label{fig:samples}
\end{figure}

\begin{figure}
  \centering
  \includegraphics[clip,trim=120 360 120 230,width=80mm]{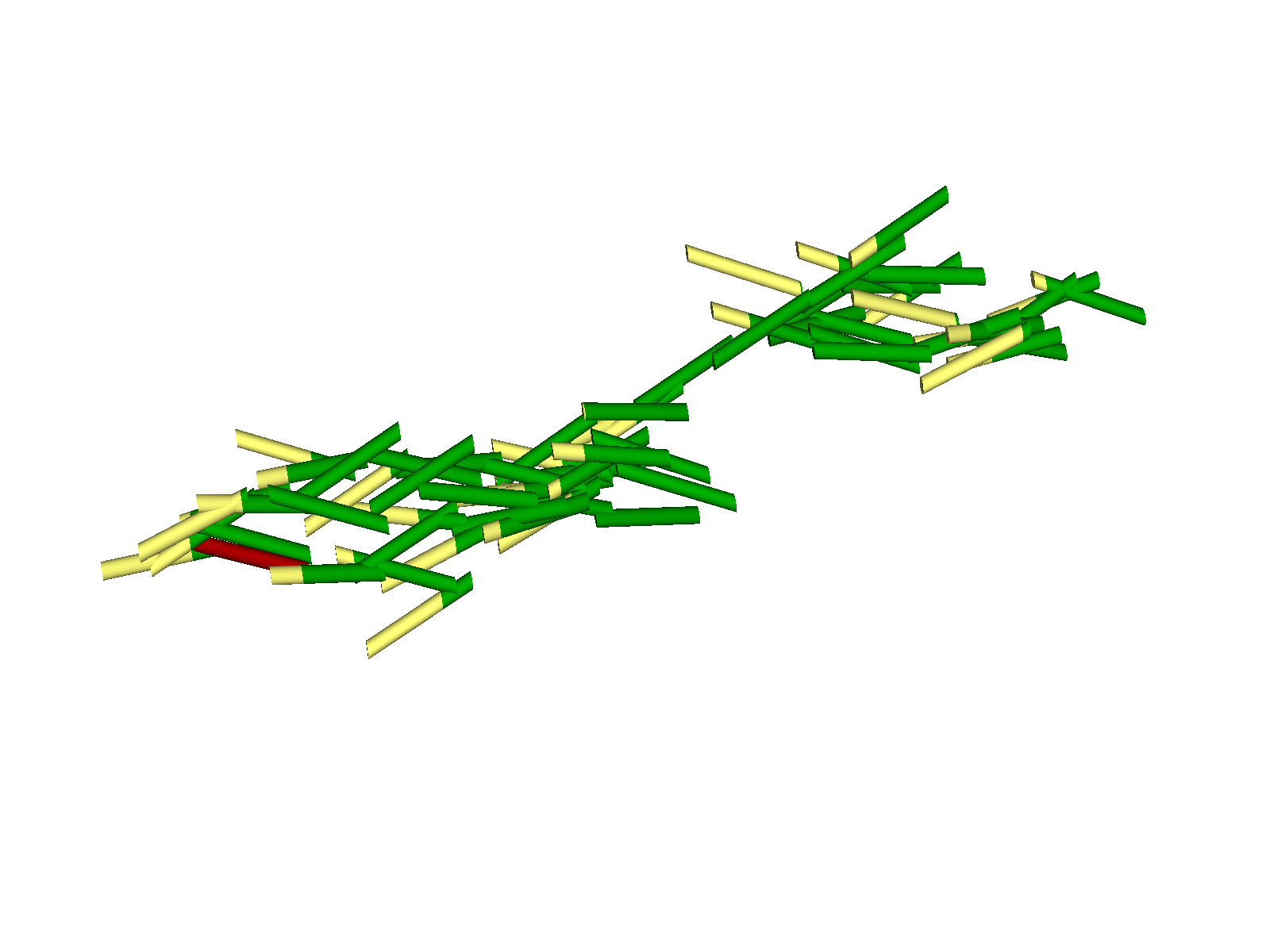}
  \caption{Sample realization with tilt $\gamma=1$ and $\eta=0.14$ %
    comprising $65$ nodes.}
  \label{fig:sample-mobile}
\end{figure}

To illustrate the directed percolation, 
consider a sample clusters depicted in Fig.~\ref{fig:samples}
obtained without mobility ($\ell=0$).
The time axis points towards the top-right corner and
without mobility all cylinders are aligned with the time axis. 
The dark red node (cylinder) indicates the source
of a message, light yellow color indicates that a node has not obtained
the message yet, and green color means that the node has it %
(at the given time).
Those nodes that never obtained the message have been omitted for clarity.
Fig.~\ref{fig:sample-mobile} illustrates the situation with mobile nodes.
From the figures we observe how the information ``flows'' only in the positive
direction of time.

When the number density of the cylinders (in space-time)
is above the critical threshold,
denoted by $n_c$, there is a positive probability that a random cylinder
belongs to an infinite cluster.
Mathematically the critical percolation density is defined as,
$$
n_c \triangleq \inf \{ n \mid \PP{ S = \infty } > 0 \}.
$$
From the information dissemination point of view, this means that there is a
positive chance that a given message 
gets distributed ``everywhere'' and does not go into extinction.

The critical number density $n_c$ obviously depends on the dimensions of the cylinders
and the amount of the movement $\ell$.
Therefore, the critical percolation threshold is often expressed using
either the critical \emph{reduced number density} $\eta_c$
or the critical \emph{volume fraction} $\phi_c$, which are both
invariant to scaling as explained earlier.
That is,
for the aligned cylinders with $\gamma=0$, 
it holds that $\eta_c$ and $\phi_c$ are independent of the dimensions
of the cylinder \cite{hyytia-cl-2012}. 
In our case, the cylinders are tilted 
to a random direction, and consequently, the criticality threshold
depends on the tilt ratio $\gamma$, and we have $\eta_c = \eta_c(\gamma)$
(i.e., $\gamma$ is a shape parameter).
Formally,
\begin{equation}\label{eq:criticality}
\eta_c(\gamma) \triangleq \inf \{ \eta \mid \PP{ S_{\gamma} = \infty } > 0 \}.
\end{equation}

\begin{figure*}
  \centering
  \small
  \setlength{\fboxsep}{1pt}
  \begin{tabular}{@{}c@{\hspace{7mm}}c@{}}
    \fbox{\includegraphics[clip,trim=140 130 150 10,height=67mm]{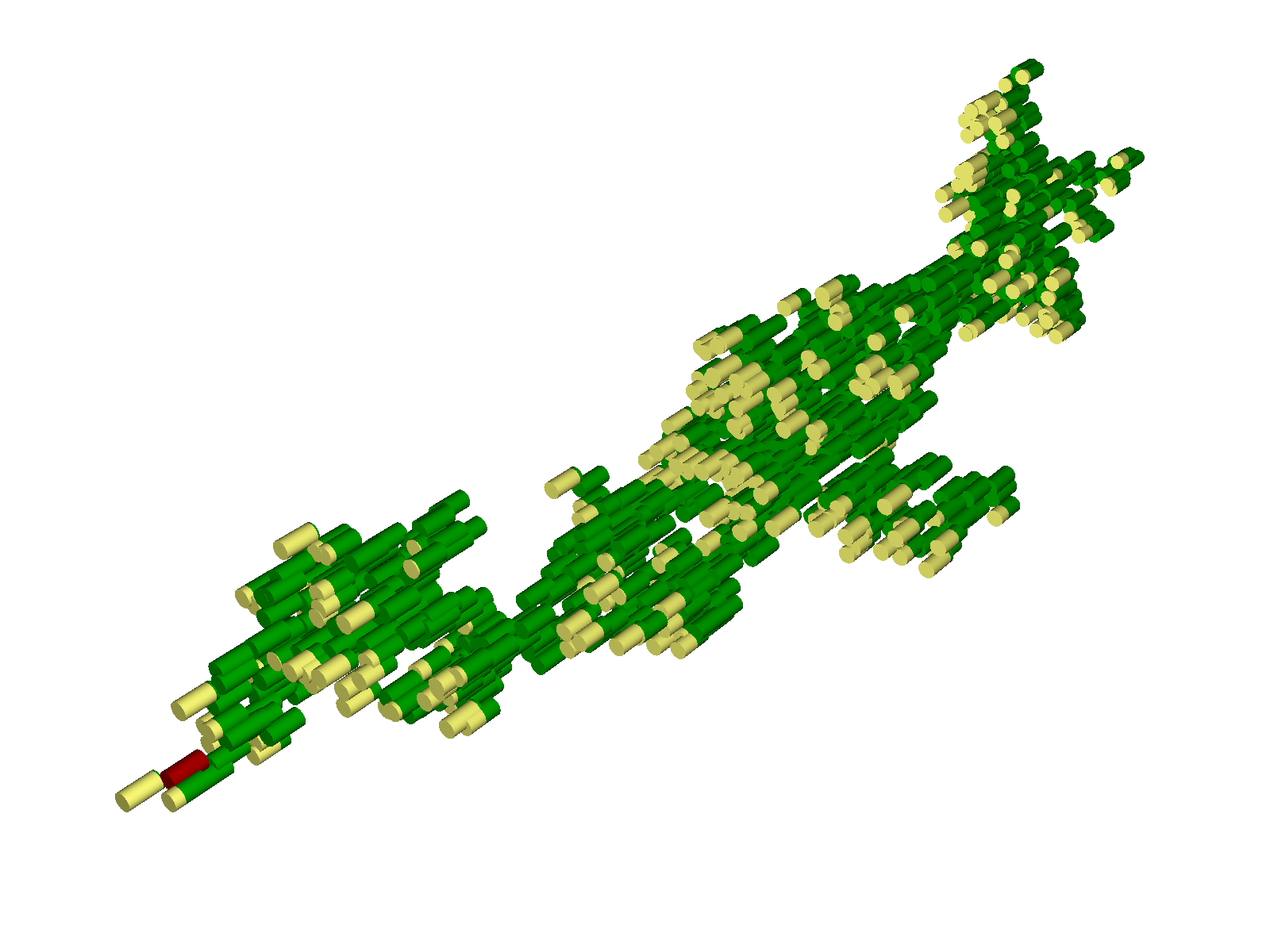}} & 
    \fbox{\includegraphics[clip,trim=175 140 100  0,height=67mm]{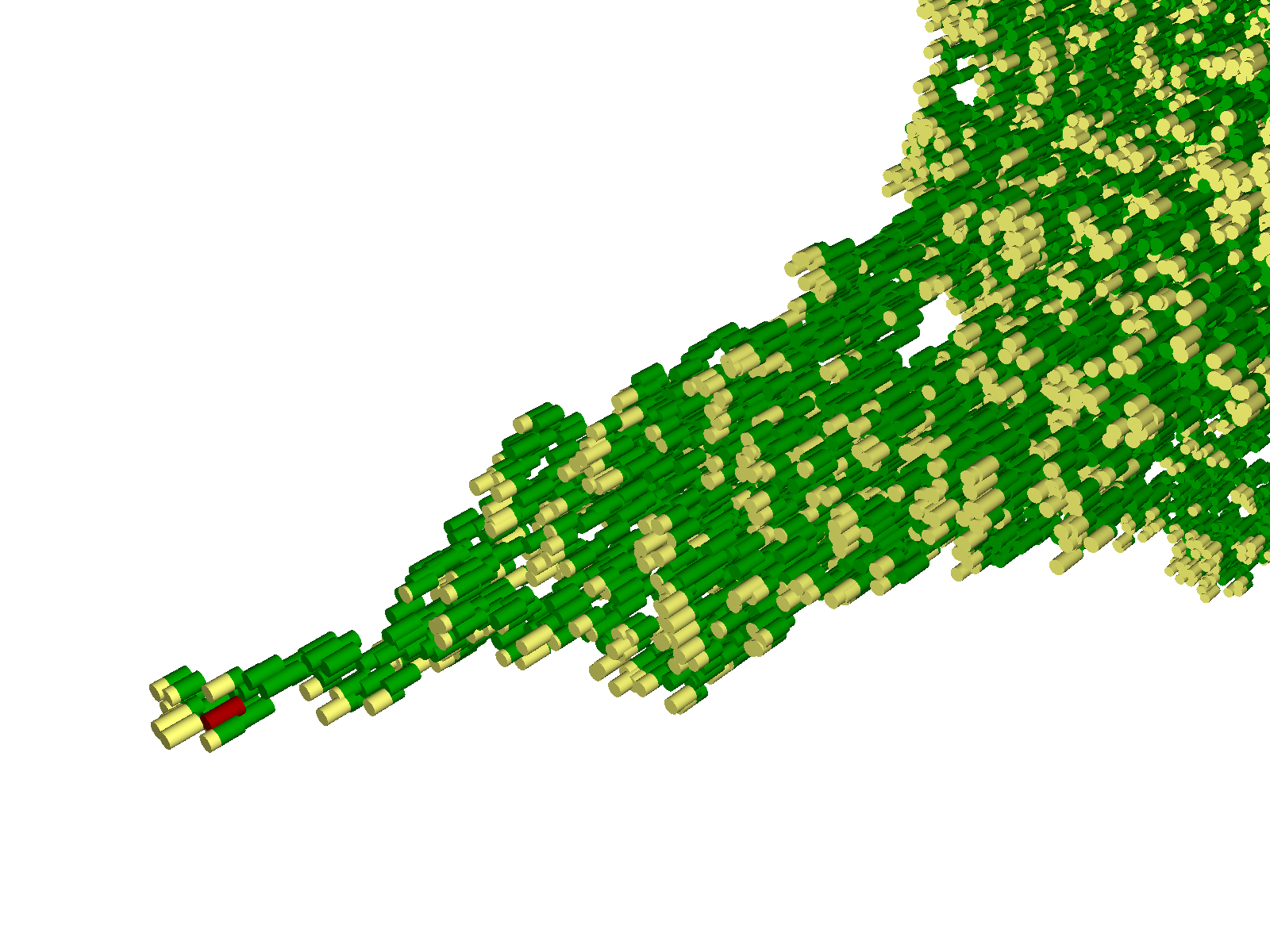}} \\[1mm]
    \small\it A sample $969$ node cluster with $\eta=0.37$. & %
    \small\it A presumably infinite cluster with $\eta=0.39$. %
  \end{tabular}
  \caption{Sample realizations of a message dissemination with
    stationary nodes. Left figure illustrates a finite cluster with $969$ nodes
    obtained with $\eta=0.37$. In the right figure, the number density is higher, $\eta=0.39$,
    which is above the percolation threshold and the sample realization (presumably) percolates.}
  \label{fig:large-samples}
\end{figure*}

Two larger sample clusters are shown in Fig.~\ref{fig:large-samples}.
The cluster in left is finite and the message goes to extinction after some time,
while the cluster on the right appears to lead to an infinite cluster, i.e.,
the system percolates implying\footnote{What happens when $\eta=\eta_c$ is not
known in general, but for $\eta<\eta_c$, by definition, no infinite cluster emerges.}
$\eta>\eta_c$.

The volume of the (tilted) cylinders is $V=\pi r^2\cdot t$
and the number density in space-time is $n=\lambda$.
Thus,
$\eta = \pi r^2 \cdot \lambda t$,
yielding
\begin{equation}\label{eq:nu-eta}
\nu = 4\,\eta.
\end{equation}
Consequently, in terms of the mean node degree,
the \emph{criticality condition states}
\begin{equation}\label{eq:criticality-nu}
  \nu_c(\gamma) = 4\,\eta_c(\gamma).
\end{equation}

An infinite cluster with respect to the directed percolation is naturally
an infinite cluster also with respect to the undirected percolation.
In \cite{hyytia-cl-2012}, we already determined the critical (undirected)
continuum percolation threshold for aligned cylinders, $\eta^*_c = 0.3312(1)$,
corresponding to the case $\ell=\gamma=0$,
which thus
serves as a strict lower bound for the directed percolation in
the same setting, $\eta_c(0)>\eta_c^*(0) = 0.3312(1)$.

When a DTN network percolates in space-time, it means that
there is a \emph{positive probability that a message makes its way to
the destination no matter how far the destination is} (in space).
In this sense, the criticality condition \eqref{eq:criticality-nu}
is a fundamental result for the transport capacity of DTN.
At the same time,
it gives no guarantees on the probability that this happens for a particular
message (cf.\ strength of the percolation), 
or on how long it takes
(which obviously depends on the distance among other things,
cf.\ \cite{jacquet-info-2010}).

\subsection{Methodology}
\label{sect:methods}

An elegant way to determine the percolation threshold is
based on the asymptotic behavior of the cluster size $S$.
In particular, the tail behaves according to
$$
\cP{S\ge s}{\eta} \sim As^{2-\tau}\, f( (\eta-\eta_c)s^\sigma ),
$$
where $\tau$ and $\sigma$ denote the so-called \emph{universal exponents}
and $A$ is some
(non-universal) constant. In three dimensions \cite{ballesteros-jphysa-1999},
\begin{equation}\label{eq:universal}
  \begin{array}{ll}
  \tau   &= 2.18906 \pm 0.00006,\\
  \sigma &= 0.4522  \pm 0.0008.
  \end{array}
\end{equation}
Near the percolation threshold, where $\eta \approx \eta_c$.
The Taylor series for $f(x)$ is $f(x)= 1 + Bx + \ldots$, which gives
$$
\cP{S\ge s}{\eta} \cdot s^{\tau-2} \sim A + AB(\eta-\eta_c)s^\sigma + \ldots,
$$
i.e., the quantity on the left-hand side becomes a constant when $\eta=\eta_c$.
The critical $\eta_c$ is then determined as follows
\cite{leath-physrevb-1976,lorenz-chemphys-2001,baker-physreve-2002,hyytia-cl-2012}.
First one collects a %
large sample set of cluster sizes 
and stores the results to bins
so that the $k$th bin, denoted by $B_k$, corresponds to
the number of samples with cluster size $s \ge 2^k$. %
For $\eta=\eta_c$, %
the quantity $B_k \cdot (2^k)^{\tau-2}$ then has a constant tail. 
Thus, by trial and error such $\eta$ is determined.

In general, there are fewer results available for the directed percolation 
than there are for the undirected percolation.
For example, the universal exponents \eqref{eq:universal} may be different.
Hence, in Section~\ref{sect:results}, we investigate the percolation
threshold of the directed model first by examining the CDF of the cluster size,
and then observe that the universal exponents may well be the same
for both cases.
To this end, we have developed a fast Monte Carlo simulation tool
that provides us with samples from the cluster size distribution.
The pseudo code of the core routine is given in Algorithm~\ref{alg:pseudo}.

\begin{figure}
  \centering
  \includegraphics[width=35mm]{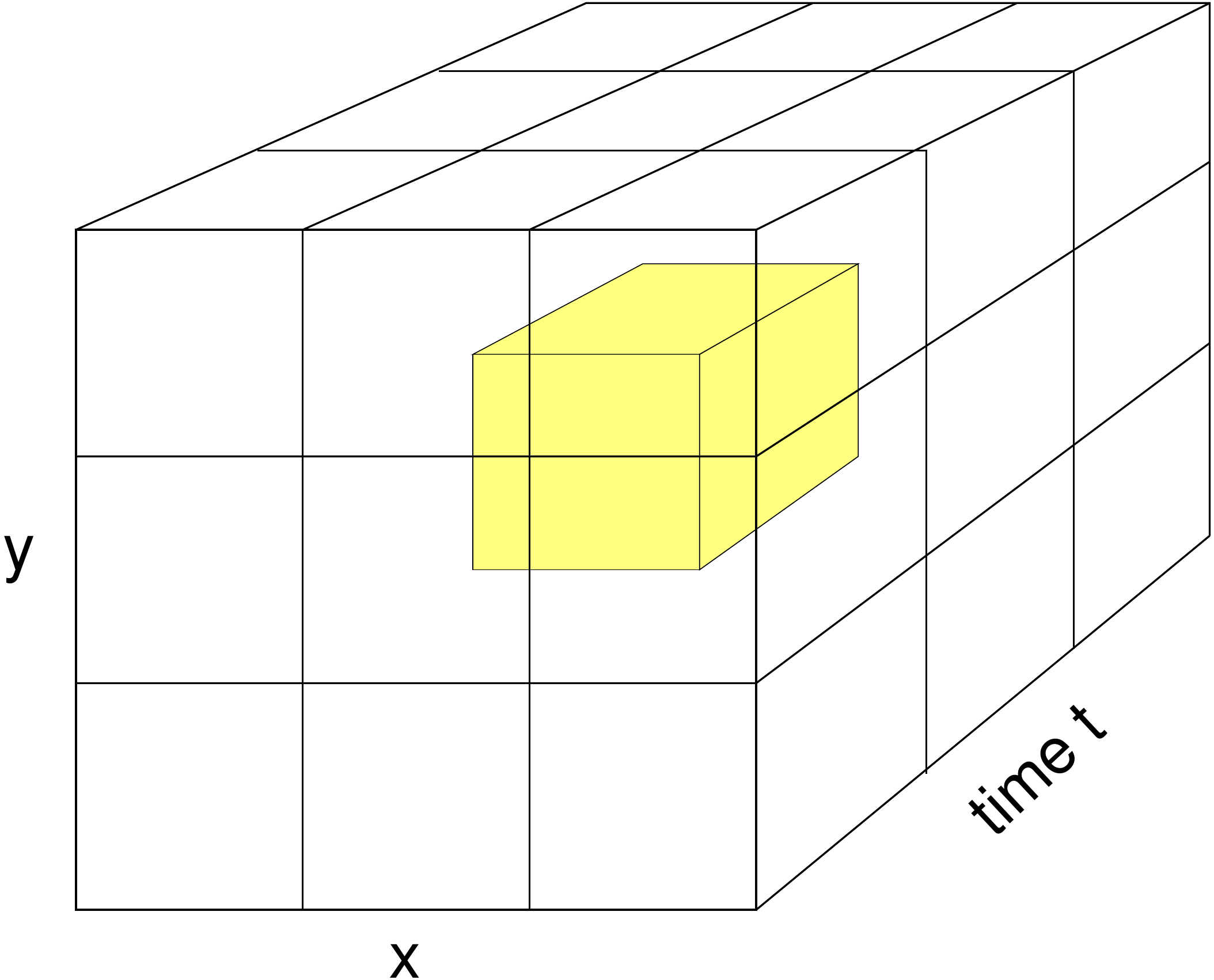}
  \caption{Partitioning of the 3D space-time.}
  \label{fig:boxes}
\end{figure}

The algorithm partitions the space-time into unit boxes as illustrated in Fig.~\ref{fig:boxes}.
The dimensions are chosen in such a way that a node appearing to the shaded box in the center
can interact during its lifetime only with nodes for which the initial 
position was in one of the $27$ surrounding boxes including the box in the center.
This gives a significant reduction in the running time of the algorithm,
where for each arriving node one needs to check the contact times with
all other nodes.
As we may be dealing with infinite clusters, an additional condition
for termination is needed. To this end, we have defined the maximum cluster size $s_{\max}$
we are interested in.

\begin{algorithm}
  \small
  {\bf iterate(cylinder $c$, time $t$)}
  \begin{algorithmic}
    \STATE Populate the $27$ boxes around $c$ (if not yet populated)
    \STATE Store the boxes to hash $\mathcal{H}$ for a fast lookup
    \FOR{each cylinder $c'$ within the neighborhood ($27$ boxes)}
    \IF{$c'$ overlaps with $c$}
      \STATE $\tau = \{z \,\mid\, z \in \text{overlap}\}$
        \COMMENT{$\tau$ is the contact time-interval}
      \IF{$\max \tau \ge t$}
        \STATE $t' = \min\{ z \in \tau\,\mid\, z\ge t\}$
        \IF{$c'$ is not marked at time $t'$}
          \STATE Schedule event $(c,c',t')$ at time $t_c$ for the pair $(c,c')$
        \ENDIF
      \ENDIF
    \ENDIF
    \ENDFOR
  \end{algorithmic}
  \vspace{1mm}
  {\bf Main routine:}
  \begin{algorithmic}
    \STATE Scale the dimensions so  that a cylinder fits inside a unit box
    \STATE Partition the 3D space to unit boxes:
    $ (x,y,z) \in B_{i,j,k}
    \;\Leftrightarrow\;
    i\le x <i+1,\; j\le y < j+1,\; k\le z < k+1.$
    \STATE Populate $B_{0,0,0}$ and store it to hash $\mathcal{H}$ for fast lookups
    \STATE Place the source node $c_0$ at $(0.5,0.5,0)$ in box $B_{0,0,0}$
    \STATE Set current time $t=0$ and cluster size $s=1$
    \STATE Set the message acquisition time of node $c_0$ to $t$
    \STATE {\bf{iterate($c_0,t$)}}
    \WHILE{scheduled events}
      \STATE Pop the next event $(c,c',t')$ from a priority queue, set $t=t'$
      \IF{$c'$ not marked {\bf and} $s<s_{\max}$}
        \STATE $s=s+1$
        \STATE Set the message acquisition time of node $c'$  to $t$
        \STATE {\bf{iterate($c',t$)}}
      \ENDIF
    \ENDWHILE
    \RETURN $s$
  \end{algorithmic}
  \caption{Pseudo code for cluster size samples.}
  \label{alg:pseudo}
\end{algorithm}

\subsection{Asymptotic behavior}

Before proceeding with the numerical results according to the percolation theory,
let us first discuss the asymptotic behavior when the distance $\ell$ is much
longer than the transmission range $d$, $\ell \gg d$.
In this case, the situation is similar to the one
analyzed in \cite{hyytia-infocom-2011,virtamo-subm-2012}: contacts with other nodes
are point contacts, which occur at a constant rate of
$$
\tilde{\lambda} \triangleq \frac{8}{\pi}\,\tilde{n}d,
$$
per unit distance on the $(x,y)$-plane.
Near the criticality threshold,
only a small fraction of nodes carry the message.
Assuming a node meets at most one node with the message,
the contact time is uniformly distributed on
interval $(0,\ell)$ along the path.
In order for the system to avoid extinction,
each node acquiring the message should 
pass it further to at least one other node on average.
That is,
$$
\frac{1}{\ell} \int_0^\ell s\,\frac{8}{\pi}\,\tilde{n}\, d\,ds > 1,
$$
giving the fluid bound (fb),
\begin{equation}\label{eq:fluid}
  \nu > \nu^{(\mathrm{fb})}_c(\gamma) \triangleq \frac{\pi^2}{4\,\gamma},
\end{equation}
which is valid when $\gamma$ tends to infinity.
At this bound, $\tilde{\lambda}\cdot\ell=2$, i.e.,
a node meets on average only two nodes before departing.
Conversely, with aid of \eqref{eq:nu-eta},
we find a \emph{scaling law} for the tail of $\eta_c(\gamma)$,
\begin{equation}\label{eq:eta-inf}
\eta_c(\gamma) \propto 1/\gamma,\quad
\text{for $\gamma \gg 1$.}
\end{equation}

It is straightforward to generalize the above 
to non-constant transitions, i.e., to the case
where the movement $\ell$ is an i.i.d.\ random variable, $\ell\sim X$,
with PDF $f(x)$. The point contact assumption
means that the transitions must be (almost surely) such that 
$\ell \gg d$.
Moreover, in a microscopic scale the movement is
assumed to be isotropic and direct, 
while in a macroscopic scale the paths can make,
e.g., smooth turns \cite{virtamo-subm-2012}.
The criticality condition is obtained as in above:
each node acquiring a message should, on average, 
pass it to one new node.
According to the theory of renewal processes (see, e.g., \cite{kleinrock-1975}), 
the PDF for transition length
$\ell$ on condition that a node obtains a message is
$$
  g(x) = x f(x) / \E{X}.
$$
(cf.\ the hitchhiker's paradox).
The remaining transition length $R$
after the acquisition %
on condition that $\ell=x$ 
is uniformly distributed on $(0,x)$ %
and $\E{R}=x/2$.
The number of new contacts $A$ in $R$ is
$$
\cE{A}{\ell=x} = \frac{4}{\pi}\, \tilde{n}\,d\,x.
$$
Hence, the mean number of contacts after 
the acquisition is
$$
\E{A} = \int_0^\infty g(x)\,\cE{A}{x} \,dx,
$$
and requiring $\E{A}>1$ then yields,
\begin{equation}
 \tilde{n}\,d > \frac{\pi\,\E{X}}{4\,\E{X^2}}.
\end{equation}
The second moment in the denominator underlines the fact
that varying transition length improves the transport capacity
in DTN communication. 

\section{Numerical results}
\label{sect:results}

\begin{figure*}
  \centering
  \begin{tabular}{cc}
    \includegraphics[clip,trim= 0 0 0 10,width=78mm]{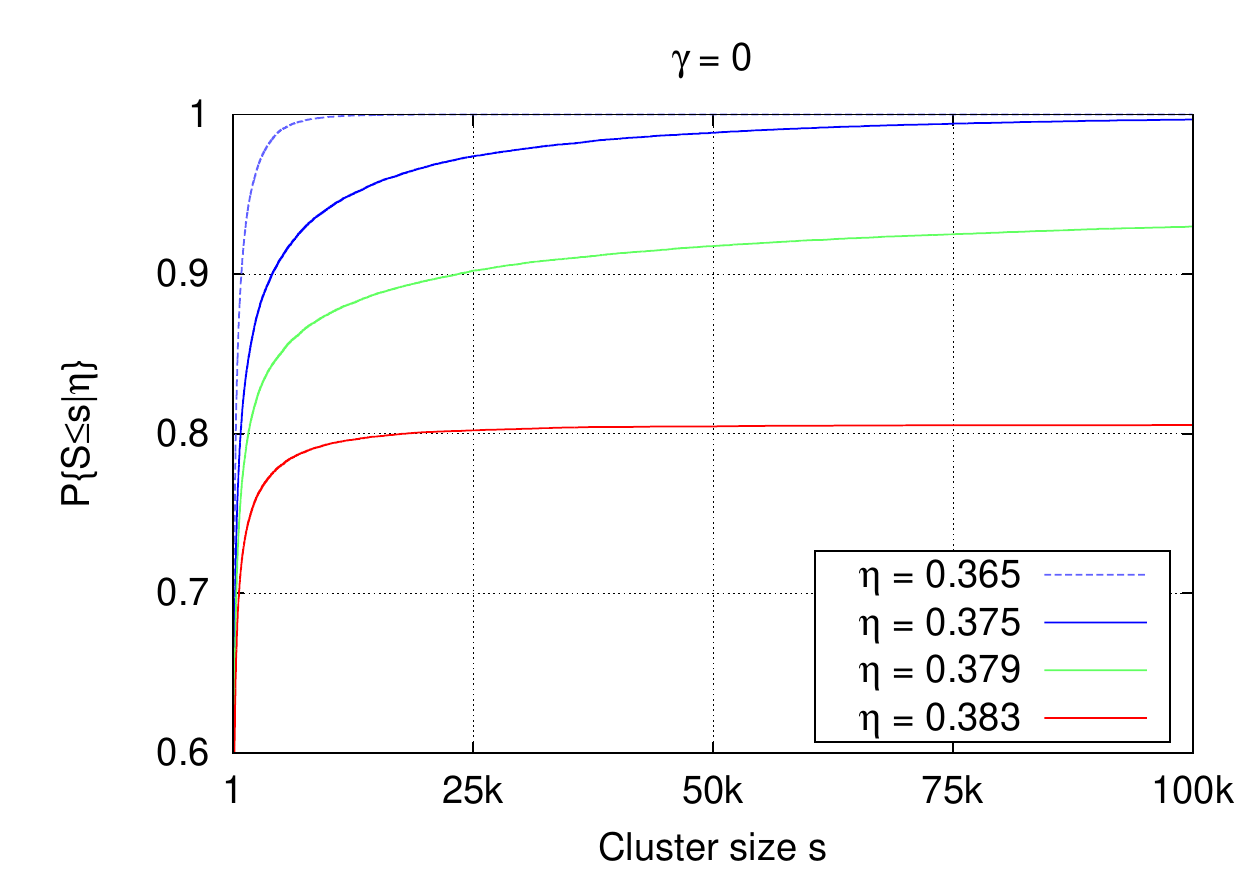} &
    \includegraphics[clip,trim= 0 0 0 10,width=78mm]{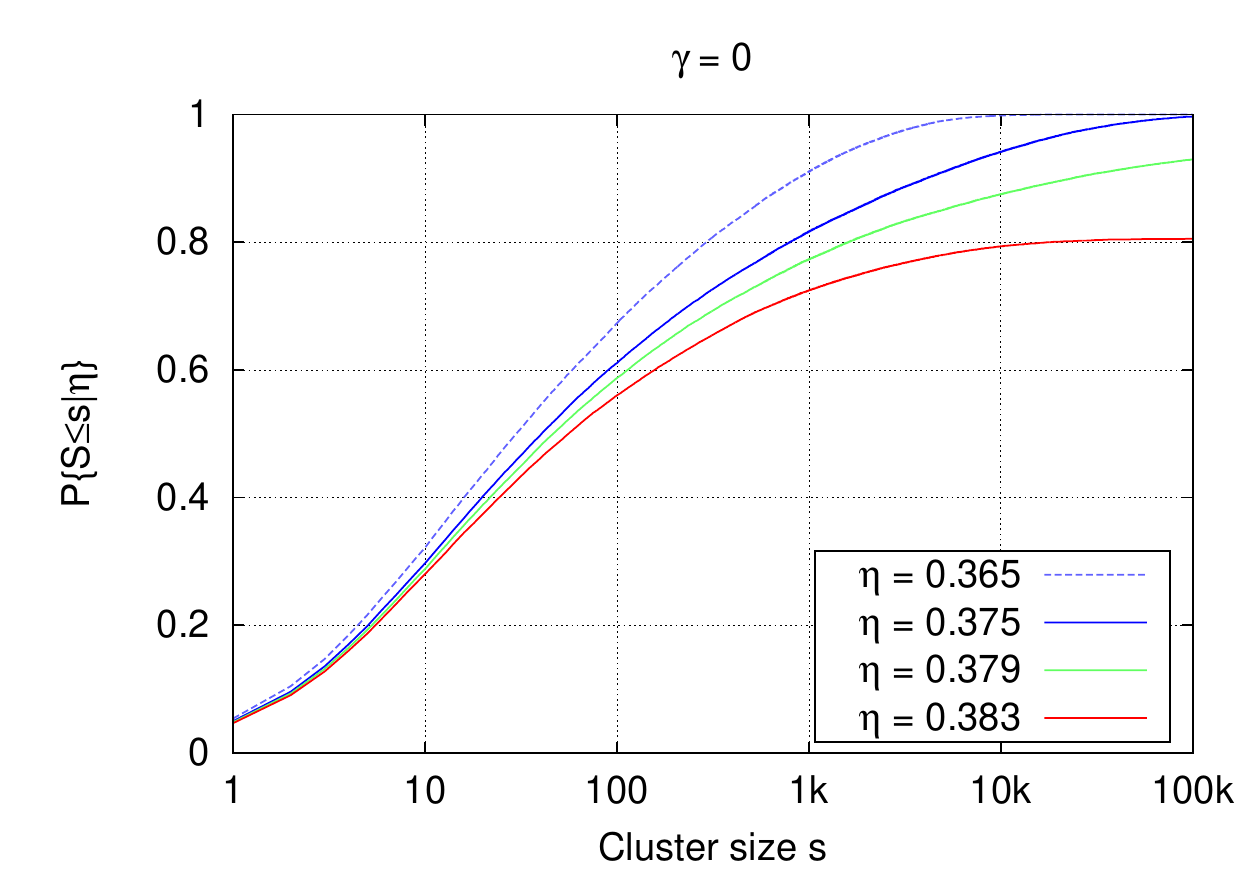}
  \end{tabular}
  \caption{Numerical results with stationary nodes, $\gamma=0$.
    Note the linear scale on the left figure and
    logarithmic scale on the right figure.}
  \label{fig:results}
\end{figure*}

\begin{figure*}
  \centering
  \begin{tabular}{cc}
    \includegraphics[clip,trim= 0 0 0 10,width=78mm]{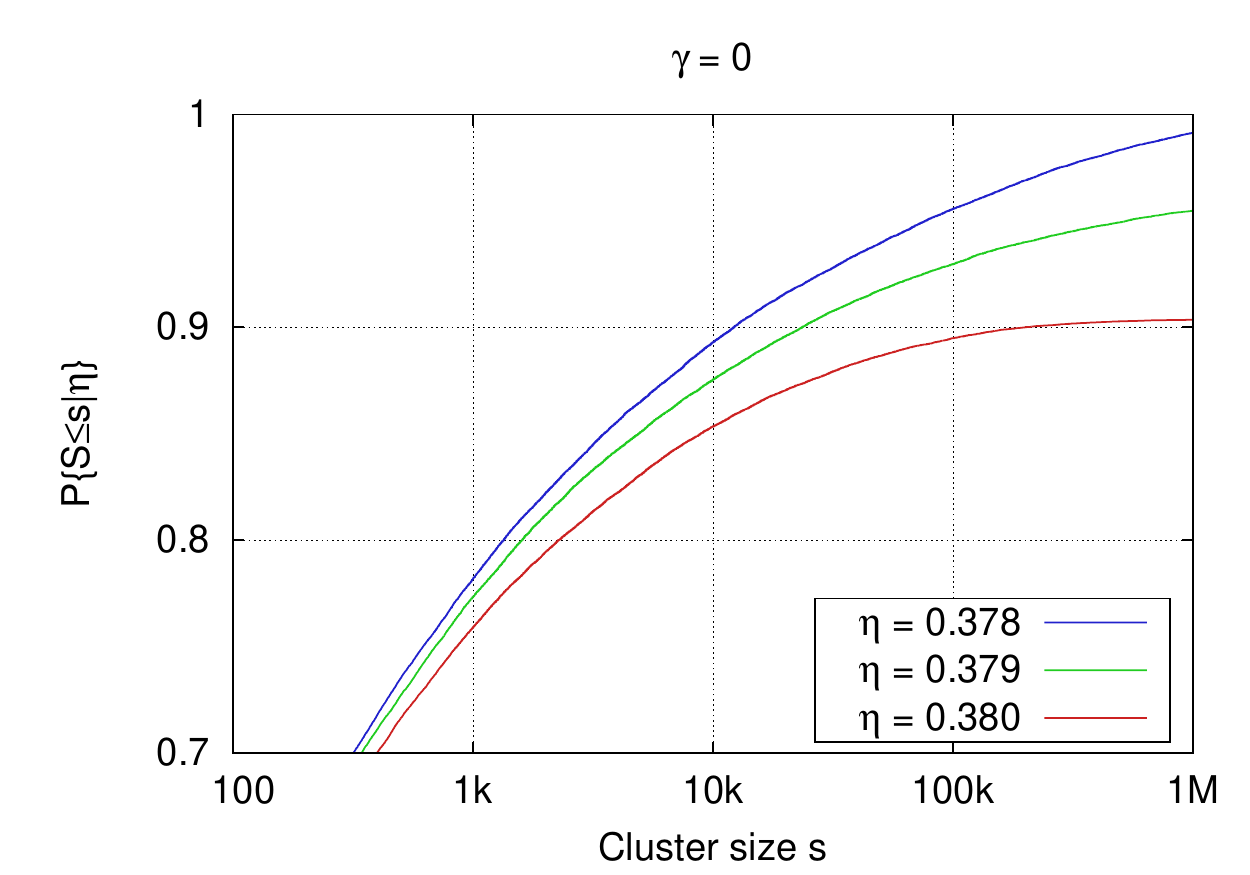} &
    \includegraphics[clip,trim= 0 0 0 10,width=78mm]{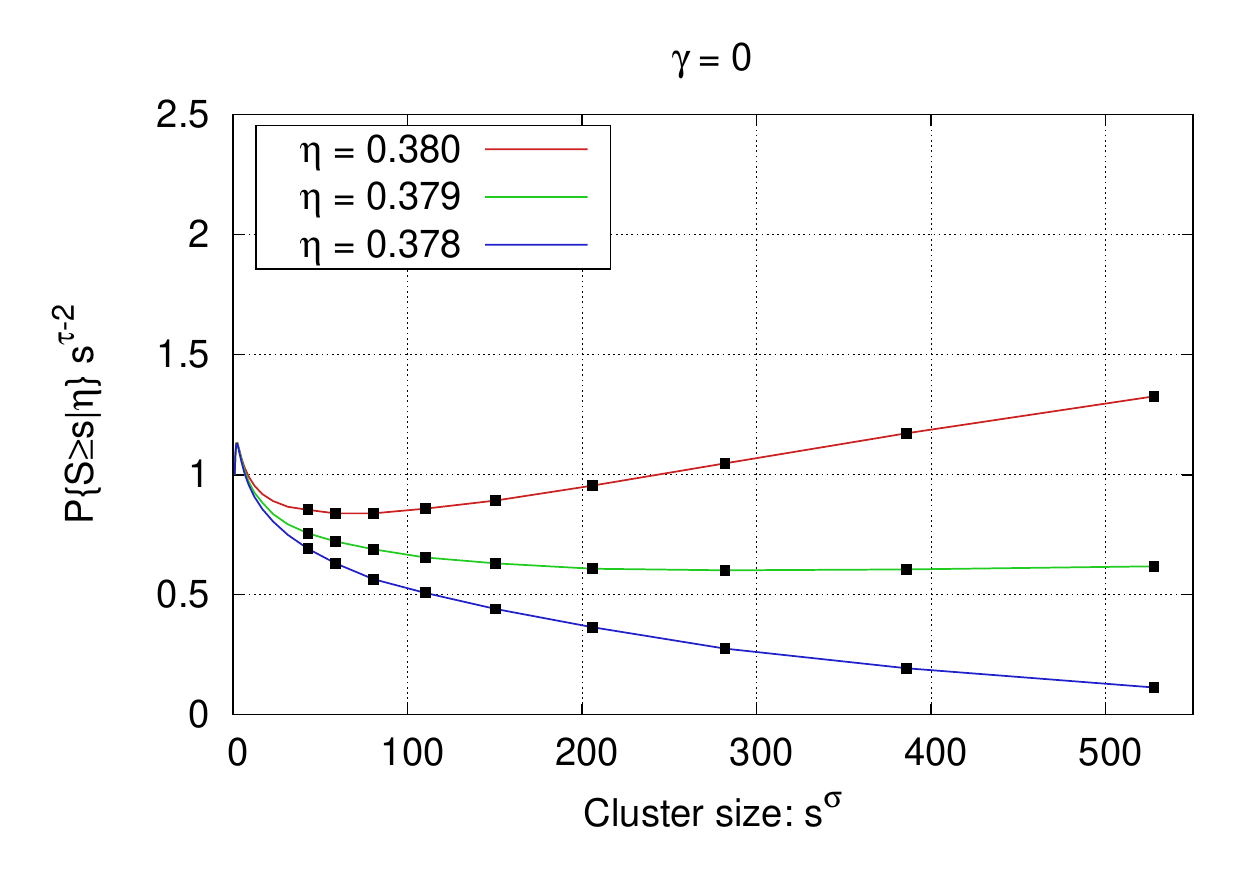}
  \end{tabular}
  \caption{Closer look at the numerical results with stationary nodes, $\gamma=0$.}
  \label{fig:results-m20}
\end{figure*}

We have carried out a large number of
Monte Carlo simulations with the aforementioned simulation tool
in order to determine numerical values for $\eta_c(\gamma)$.
The sample size for each $(\eta,\gamma)$-pair is $60000$ clusters.
The parameter $s_{\max}$, defining the maximum cluster size, %
was chosen to be $2^{20}+1 = 1048577$, i.e., about $1$ million.
With these values, the running time is still 
reasonably short with a standard PC.
Such clusters are rather large creatures, e.g.,
there are more than $10^{12}$ node pairs,
and a na\"{\i}ve implementation would choke on them.

\subsection{Stationary nodes}
\begin{figure}
  \centering
  \begin{tabular}{@{}c@{\hspace{2mm}}c@{}}
    \includegraphics[width=41mm]{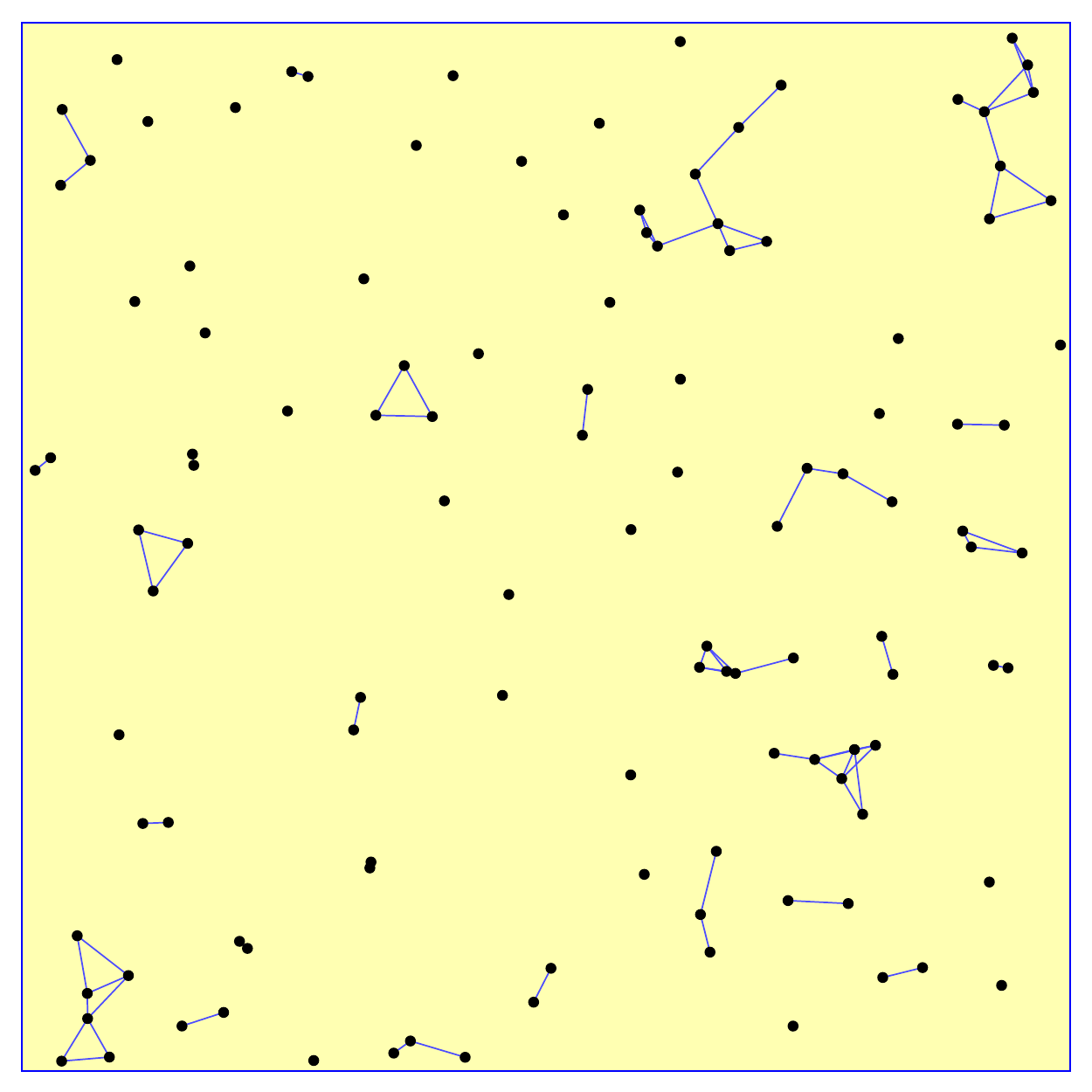} &
    \includegraphics[width=41mm]{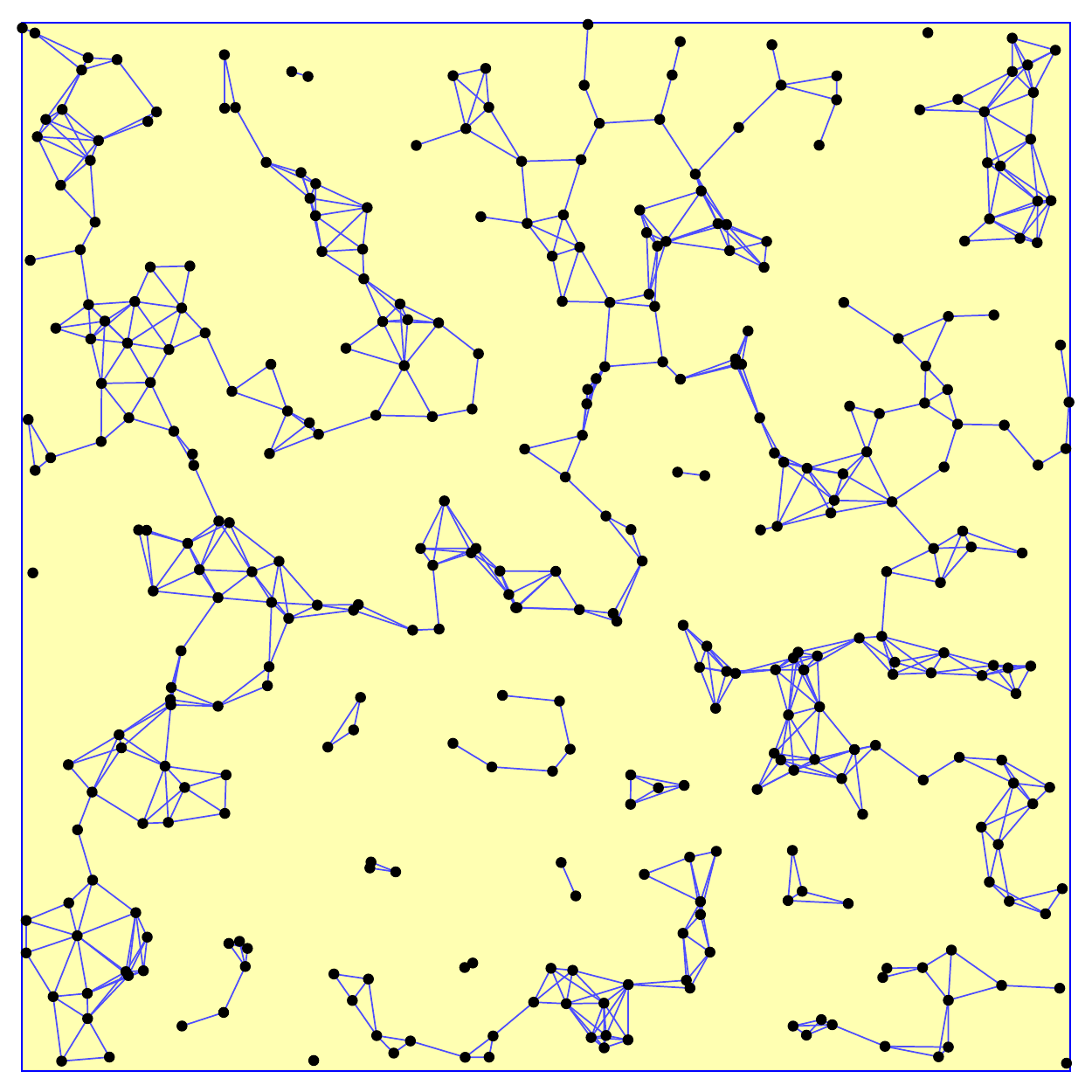}\\
    \scriptsize $\nu=1.52$ &
    \scriptsize $\nu=4.51$
  \end{tabular}
  \caption{Snapshot of two networks.}
  \label{fig:sample-nets}
\end{figure}

First we assume no mobility, i.e., $\gamma=0$, and
vary the reduced number density, $\eta=0.365,\ldots,0.383$,
which corresponds to the mean node degree of 
$\nu = 1.46, \ldots, 1.53$.
If $\nu \approx 4.51$ or higher, Gilbert's disc model percolates and
a gigantic component emerges (at any given time instant) enabling
multi-hop communication \cite{franceschetti-info-theory-2007}.
However, in our case the node density is clearly below that and
one has to resort to DTN-style communication.
This is illustrated in Fig.~\ref{fig:sample-nets}, where
$\nu=1.52$ in the left figure and $\nu=4.51$ in the right figure.

Recall that the %
percolation threshold $\eta_c$ corresponds
to the smallest node density at which a gigantic component emerges,
$$
\eta_c = \inf \{ \eta \mid \PP{ S = \infty } > 0 \}.
$$
Therefore, if CDF of the cluster size $S$
remains asymptotically below $1$,
$\lim_{s\to\infty} \PP{S<s} = p$ and $p<1$, then
the given system is above the percolation threshold.
The quantity $1-p$, referred to as the \emph{strength},
is the probability that a cluster starting from a random node is infinite
\cite{stauffer-1994}.

The simulation results are illustrated in Fig.~\ref{fig:results}.
The left figure depicts CDF with a linear scale on the $x$-axis, 
and the right figure with a logarithmic scale.
The logarithmic scale gives more insight to the behavior.
The CDF for $\eta=0.365$ converges to $1$ at $s \approx 10^4$,
i.e., in practice every realization goes to extinction 
and no message will be delivered for more than to about $10^4$ nodes.
However, the CDF for $\eta=0.379$ seems to converge to a finite value less than $1$,
suggesting that a small fraction of messages may survive forever.
Increasing $\eta$ further to $0.383$ improves the chances 
of %
hitting a gigantic cluster to about $20\%$, 
i.e., with $\eta=0.383$ the strength of the percolation is about $0.2$.

Conservatively, we observe that $\eta=0.383$ is above the percolation threshold,
while $\eta=0.375$ is most likely below it, $0.375< \eta_c < 0.383$.
Comparing the directed percolation to the undirected one,
we see that the direction constraint causes $\eta_c$ to increase from
$0.3312(1)$ to about $0.38$, i.e., about $15\%$ increment.
This is in agreement with the observations made in \cite{hyytia-cl-2012},
where an actual content dissemination system (obeying the causality)
was also evaluated by means of simulations.

\begin{figure*}
  \centering
  \begin{tabular}{cc}
    \includegraphics[clip,trim= 0 0 0 10,width=78mm]{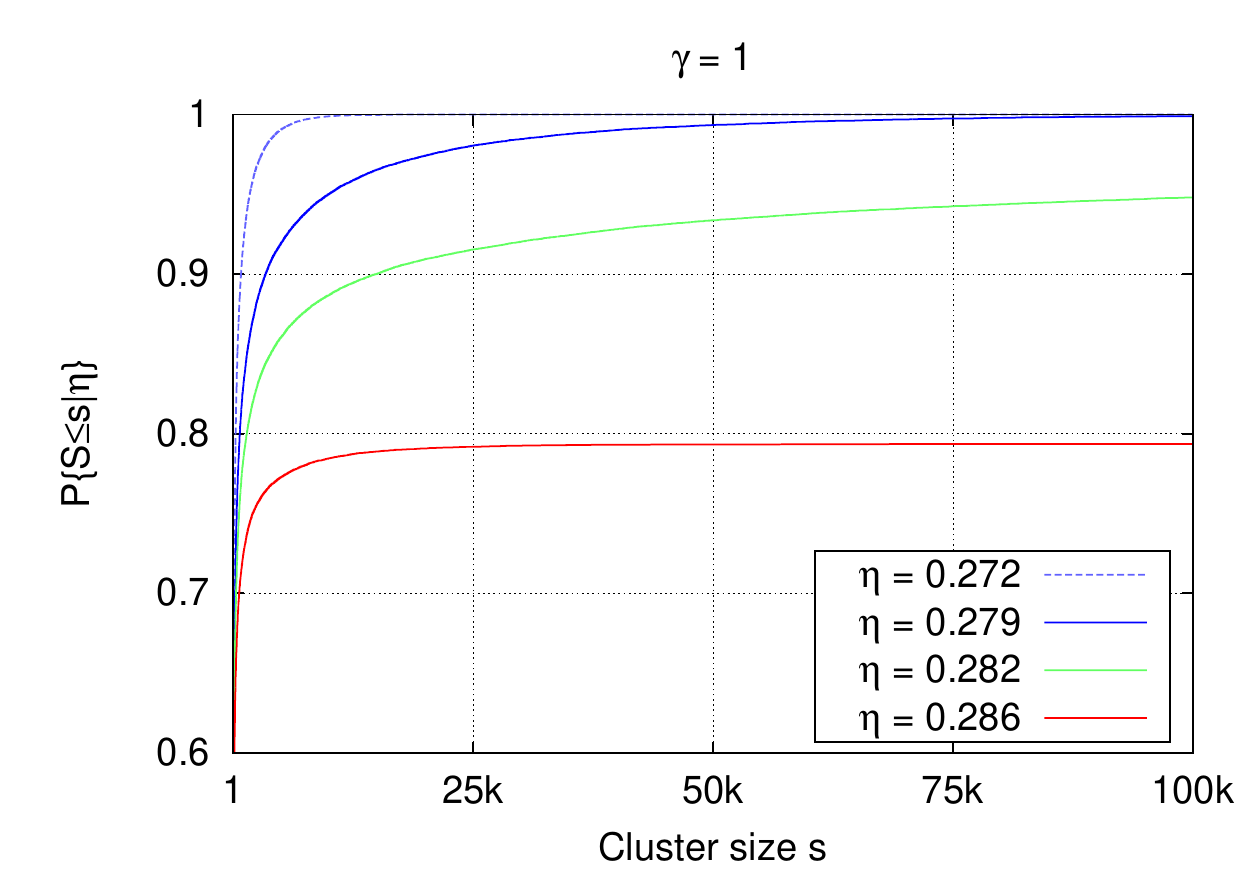} &
    \includegraphics[clip,trim= 0 0 0 10,width=78mm]{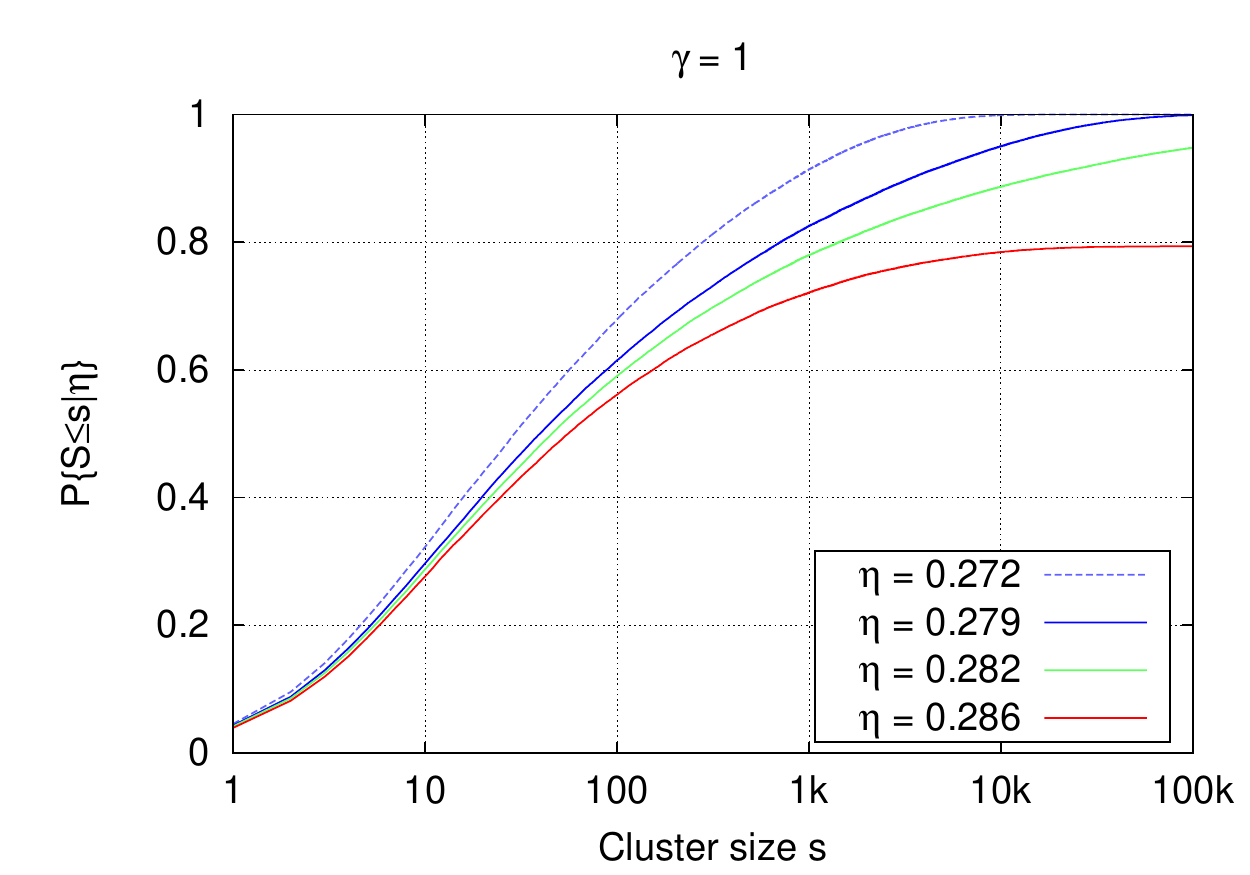}
  \end{tabular}
  \caption{Numerical results with mobile nodes, $\gamma=1$.}
  \label{fig:results-1}
\end{figure*}

\begin{figure*}
  \centering
  \begin{tabular}{cc}
    \includegraphics[clip,trim= 0 0 0 10,width=78mm]{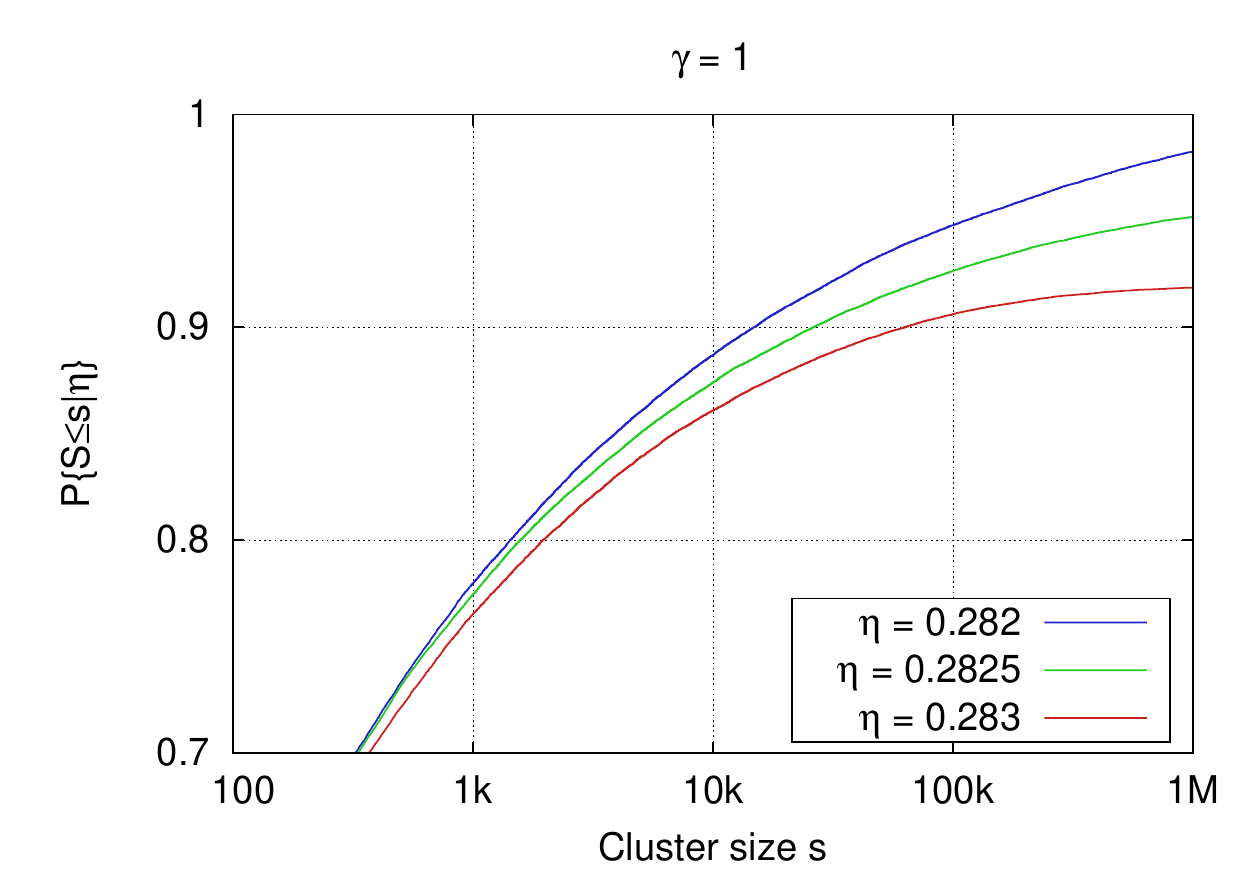} &
    \includegraphics[clip,trim= 0 0 0 10,width=78mm]{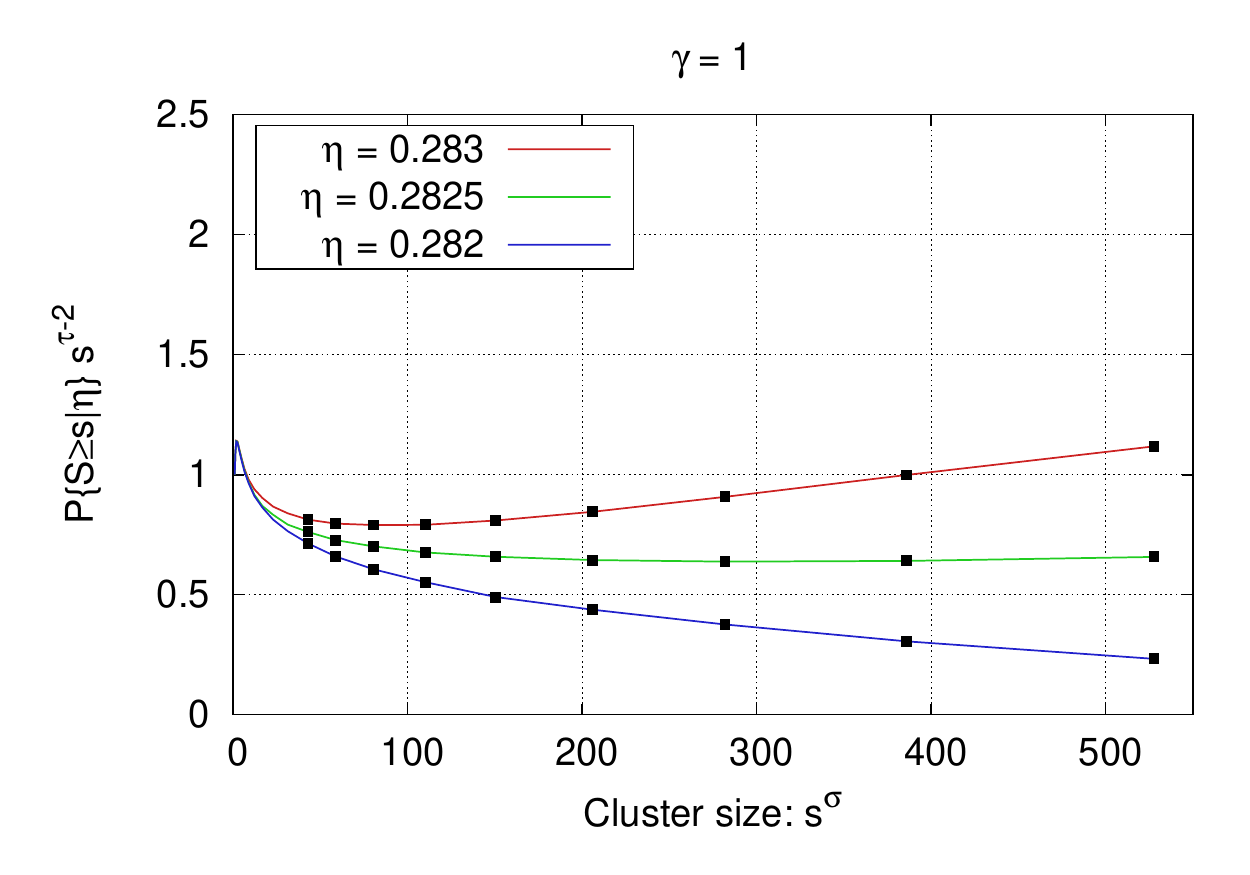}
  \end{tabular}
  \caption{Closer look at the results with mobile nodes, $\gamma=1$.}
  \label{fig:results-1b}
\end{figure*}

\subsubsection{Universal exponents}
As mentioned, it is not clear whether the universal exponents
\eqref{eq:universal} are the same for the directed model and
the normal undirected model.
Fig.~\ref{fig:results-m20} (left)
shows CDF of the cluster size distribution 
in the interesting region near the percolation threshold.
The graph clearly \emph{suggests} that $0.378< \eta_c(0) < 0.380$
and our estimate is $\eta_c(0) \approx 0.379$.
In Fig.~\ref{fig:results-m20} (right), 
the universal exponents according to \eqref{eq:universal} are used.
The curve corresponding to $\eta=0.379$ has a constant tail, 
which 
suggests that the universal exponents are the same for the undirected
and directed percolation models, 
and we assume so in the rest of the paper.
(cf.\ also the Appendix).

\subsection{Mobile nodes}

Next we set parameter $\gamma$ to $1$, i.e., each node moves a distance equal to
the transmission range before departing.
The numerical results are given in Fig.~\ref{fig:results-1},
where the left figure is again in linear scale and the right in logarithmic.
We observe that the critical $\eta_c(1) \approx 0.282$,
as the curve with $\eta=0.279$ clearly converges to $1$ and 
the curve with $\eta=0.286$ stabilizes at about $0.8$.
Fig.~\ref{fig:results-1b} takes a closer look at the interesting regime. 
Left figure depicts the CDF, which suggests that $0.282 < \eta_c(1) < 0. 283$.
In the right figure, we again use the universal exponents
and the curve with $\eta=0.2825$ indeed appears to have a constant tail.
However, a closer inspection of the curve reveals that in fact it
is slightly increasing already at $s^\sigma=500$,
suggesting $\eta_c < 0.2825$.
As three digits is more than enough for our purposes,
we conclude that $\eta_c(1) \approx 0.282$.

Recall that $\nu=4\,\eta$ according to \eqref{eq:nu-eta}, i.e.,
the minimum feasible mean node degree $\nu_c(\gamma)$
is obtained by multiplying the corresponding $\eta_c(\gamma)$ by $4$.
According to our numerical results,
$\eta_c(1) \approx 0.75\cdot\eta_c(0)$, i.e., $25\%$ smaller.
This means that both the node density $n$
and the mean node degree $\nu$ can also be $25\%$ smaller,
i.e., even a small mobility equal to the transmission range 
improves the message forwarding capacity considerably (in space-time).

\begin{figure}
  \centering
  \includegraphics[width=78mm]{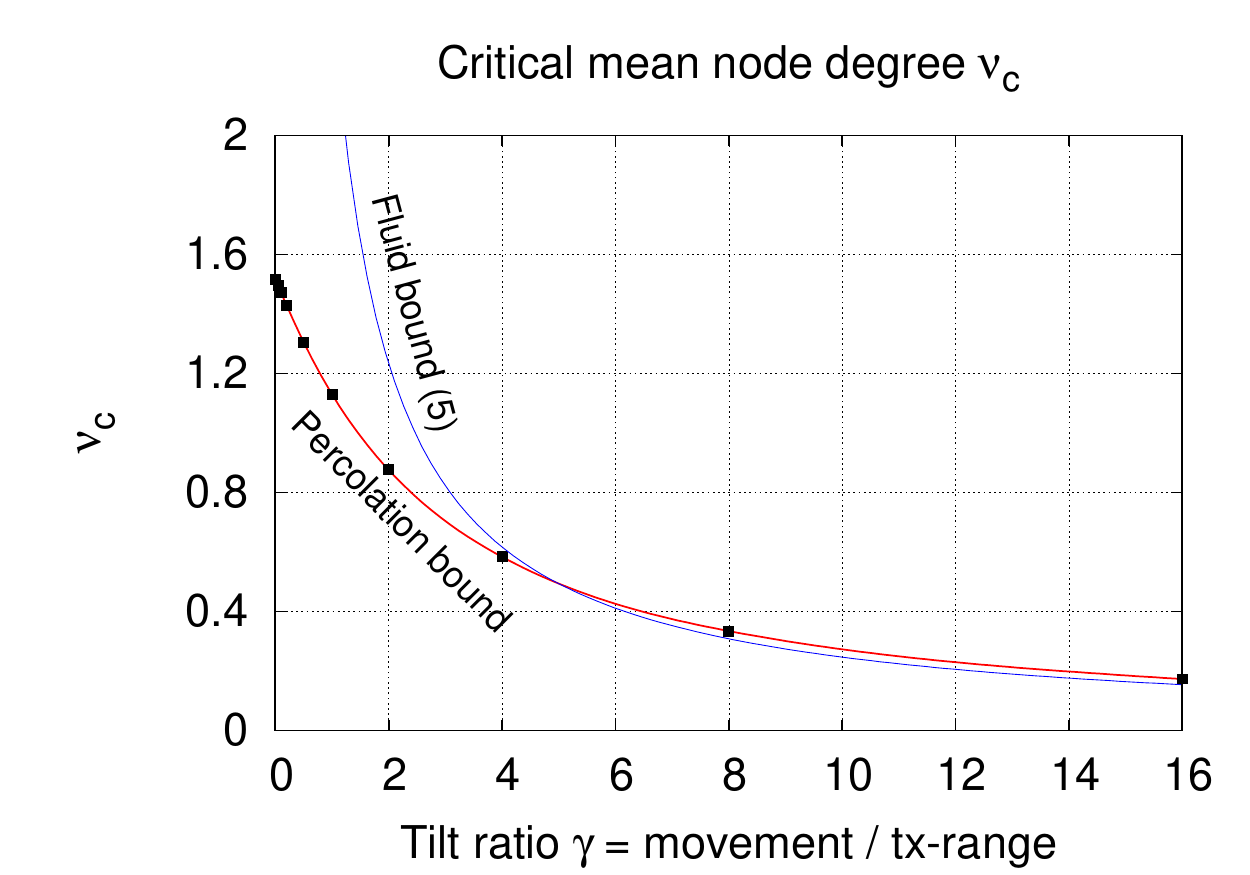}
  \caption{Critical mean node degree $\nu_c(\gamma)$ as a function of the tilt ratio $\gamma$
    characterizing the relative mobility in the model.
    The fluid bound curve is according to \eqref{eq:fluid} and valid
    at the limit $\gamma \to \infty$. 
    The other curve is due to percolation theory,
    $\nu_c(\gamma) = 4\, \eta_c(\gamma)$.}
  \label{fig:nu}
\end{figure}

The critical reduced number density $\eta_c(\gamma)$ for
several other values of $\gamma$ is determined in the Appendix.
Fig.~\ref{fig:nu} depicts the behavior of 
the critical mean node degree $\nu_c(\gamma)$
as a function of $\gamma$.
The \emph{fluid bound} curve is according to \eqref{eq:fluid} and valid
at the limit $\gamma \to \infty$. 
The other curve corresponds to the
critical percolation threshold, $\nu_c(\gamma) = 4\, \eta_c(\gamma)$,
the values of which we have obtained numerically in this paper.
We observe that the highest gain
from an additional movement of $\Delta \ell$ is
obtained when $\gamma=0$, as expected.

The two curves differ initially, where \eqref{eq:fluid} is not valid.
However, already at $\gamma \approx 4$ the two curves are almost matching.
Consequently, for a small movement one needs to analyze the situation
according to the percolation, but as soon as, say, $\gamma \ge 4$,
the fluid flow analysis already gives accurate results.

\section{Conclusions}
\label{sect:conclusions}

DTN is designed to operate in settings where connectivity is intermittent at best.
In this paper, we have analyzed necessary conditions for DTN type communication.
To this end, we assumed an elementary mobility model where nodes join the network
for a constant time $t$ and then depart. 
During the sojourn time, they move a distance
$\ell$ to a random direction.
We assume a large network, where the ability to sustain a message 
indefinitely means that the message will eventually also
reach its final destination(s).
The stochastic model of DTN is amenable to analysis by means of the percolation theory.
In particular, we studied the so-called directed continuum percolation
in space-time, where the objects are cylinders 
with height $t$ and the diameter equal to the transmission range $d$
sheared (tilted) around the time axis by an amount corresponding to the movement $\ell$.
We showed that the critical percolation threshold
in terms of the critical reduced number
density $\eta_c$ depends only on the tilt ratio $\gamma=\ell/d$,
$\eta_c=\eta_c(\gamma)$,
where $\ell$ denotes the distance nodes move and
$d$ is the transmission range.
Numerical values for $\eta_c(\gamma)$ were obtained by
means of Monte Carlo simulations, where universal exponents 
of the undirected model were utilized.
Some evidence to support the hypothesis that the universal exponents
are the same for the directed model was also given.
Moreover, 
the asymptotic behavior of $\eta_c(\gamma)$ when $\gamma$ 
tends to infinity was derived.
In terms of the mean node degree $\nu$, our main result states
that $\nu > 4\,\eta_c(\gamma)$ in order for a large DTN network
to be operational.
A general observation is that
increased mobility improves the transport capacity of DTN %
and allows a lower node density, as expected.

The elementary mobility model, as well as, the symmetric Gilbert's disc communication
model, represent, in some sense, a strict lower bound for more realistic scenarios.
That is, if, e.g., the sojourn time $t$ or the distance $\ell$ were random
variables, or, if directional antennas were used, then the shapes and
the orientation of the objects would vary more, increasing ``disorder''
and yielding a lower critical percolation threshold.
Such studies are left as a future work.

\newcommand{\kuvapari}[3]{%
  \begin{figure*}
    \centering
      \includegraphics[width=65mm]{#1-cdf} \hspace{10mm} %
      \includegraphics[width=65mm]{#1-bin}
    \caption{Numerical results with mobile nodes corresponding to tilting
      cylinders with $\gamma=#2$.}
    \label{fig:results-#3}
  \end{figure*}}

\newlength{\triplalena}
\newlength{\triplalenb}
\setlength{\triplalena}{35mm}
\setlength{\triplalenb}{33mm}
\newcommand{\tripla}[6]{%
  \begin{figure*}
    \centering
    \begin{tabular}{c@{}c@{}c@{}c}
      \includegraphics[clip,trim= 0 0 0 10,height=\triplalena]{{m20-#1-cdf-zoom}.pdf} &
      \includegraphics[clip,trim=61 0 0 10,height=\triplalena]{{m20-#2-cdf-zoom}.pdf} &
      \includegraphics[clip,trim=61 0 0 10,height=\triplalena]{{m20-#3-cdf-zoom}.pdf} &
      \includegraphics[clip,trim=61 0 0 10,height=\triplalena]{{m20-#4-cdf-zoom}.pdf} \\
      \includegraphics[clip,trim= 0 0 0 23,height=\triplalenb]{{m20-#1-bin}.pdf} &
      \includegraphics[clip,trim=61 0 0 23,height=\triplalenb]{{m20-#2-bin}.pdf} &
      \includegraphics[clip,trim=61 0 0 23,height=\triplalenb]{{m20-#3-bin}.pdf} &
      \includegraphics[clip,trim=61 0 0 23,height=\triplalenb]{{m20-#4-bin}.pdf}
    \end{tabular}
    \caption{Numerical results for $\eta_c(\gamma)$ with $\gamma=#5$.}
    \label{fig:results-#6}
  \end{figure*}}

\tripla{0.05}{0.1}{0.2}{0.5}{0.05, 0.1, 0.2, 0.5}{01-05}
\tripla{2}{4}{8}{16}{2, 4, 8, 16}{2-16}

\newcommand{\SortNoop}[1]{}

\appendix

\begin{figure}
  \centering
  \includegraphics[clip,trim=11  4 11 6,height=30mm]{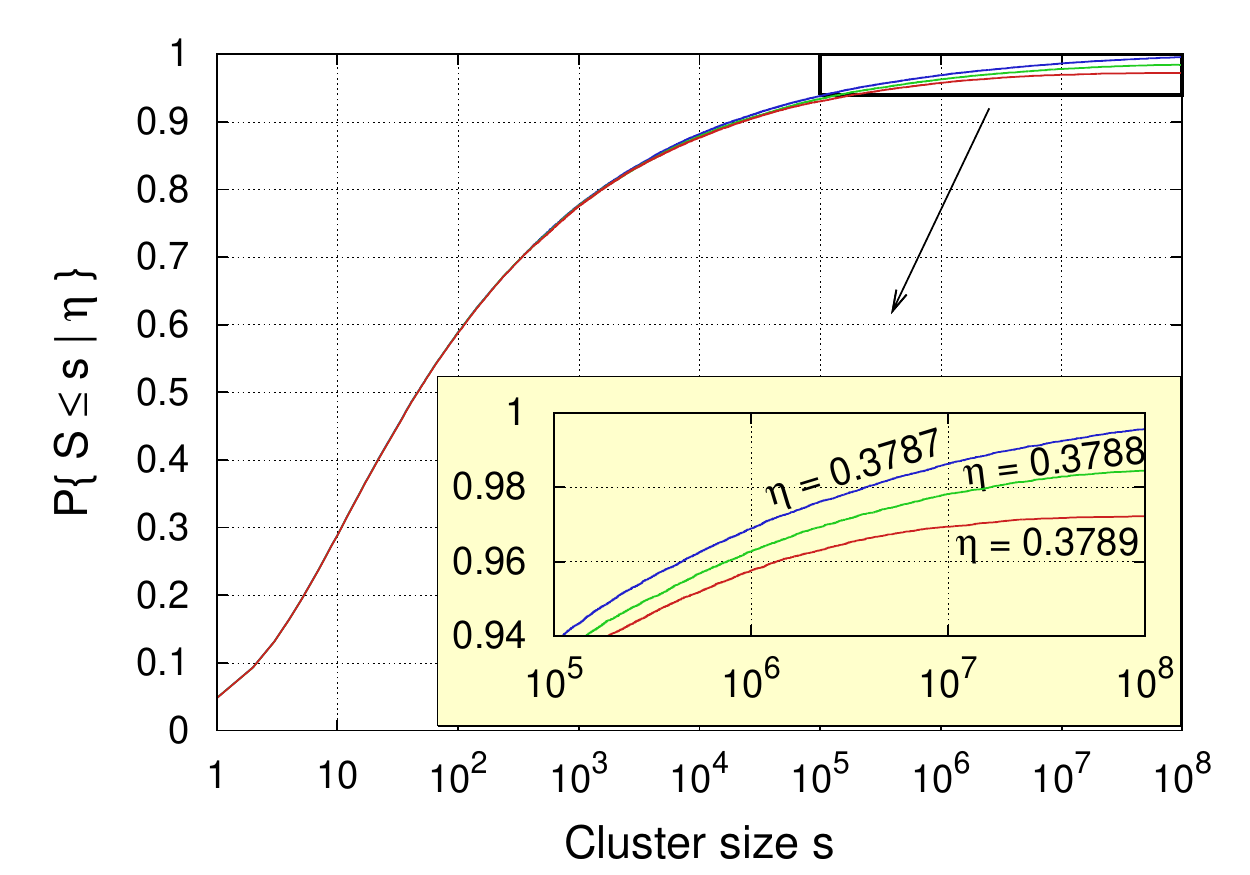}
  \includegraphics[clip,trim=11 16 19 6,height=30mm]{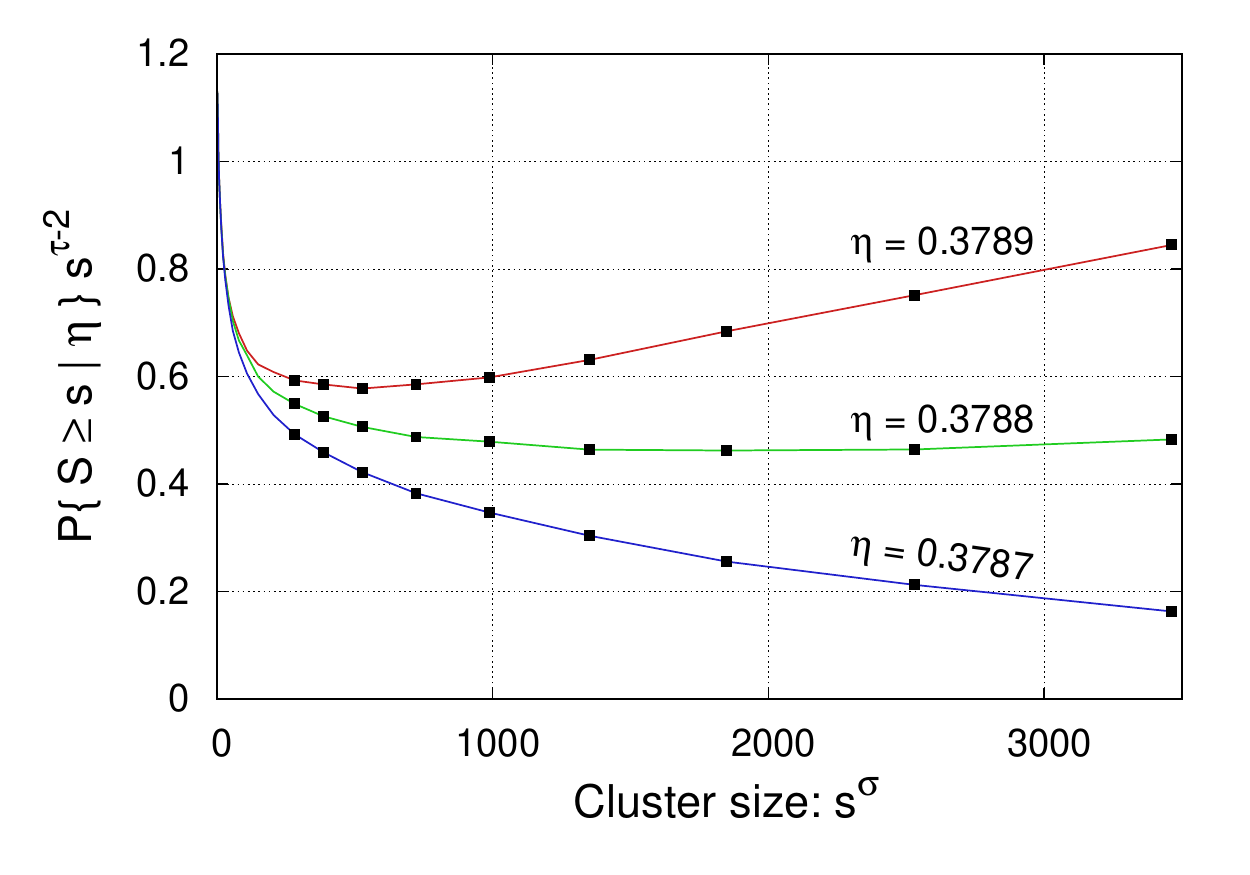}
  \caption{Results with $\gamma=0$ and $s_{\max}=10^8$, i.e., $100$ million.}
  \label{fig:100M}
\end{figure}

An %
estimate for the critical reduced number density
$\eta_c(\gamma)$ is determined
for $\gamma{=}0.05,\, 0.1,\, 0.2,\, 0.5,\, 2,\, 4,\, 8,\, 16$
in Figs.~\ref{fig:results-01-05}-\ref{fig:results-2-16}.
In each case, parameter $\eta$ obtains values from both sides
of the corresponding critical percolation threshold.
Table~\ref{tbl:results} summarizes the results. %
In Fig.~\ref{fig:100M},  the maximum cluster size is very large,
$s_{\max}=10^8$, suggesting
$\eta_c(0) = 0.3788(1)$ and also supporting
the hypothesis on the universal exponents.

\begin{table}
  \centering
  \begin{tabular}{|r@{}ll|r@{}ll|}
    \hline
    \multicolumn{2}{|c}{$\gamma$} & \multicolumn{1}{c|}{$\eta_c(\gamma)$} &
    \multicolumn{2}{|c}{$\gamma$} & \multicolumn{1}{c|}{$\eta_c(\gamma)$}\\
    \hline
       &&&&&\\[-2.9mm]
    $0$&$.0$  & $0.3788$ & $1$&  & $0.2825$\\
    $0$&$.05$ & $0.3735$ & $2$&  & $0.2189$\\
    $0$&$.1$  & $0.368$  & $4$&  & $0.1459$\\
    $0$&$.2$  & $0.357$  & $8$&  & $0.0836$\\
    $0$&$.5$  & $0.3262$ & $16$& & $0.04325$\\
    \hline
  \end{tabular}
  \caption{Critical reduced number density $\eta_c(\gamma)$ based on the
    Monte Carlo simulations, where $\eta_c(8) \approx 2\,\eta_c(16)$
    in agreement with \eqref{eq:eta-inf}.}
  \label{tbl:results}
\end{table}

\end{document}